\documentclass[a4paper,11p]{article}
\pdfoutput=1 % if your are submitting a pdflatex (i.e. if you have
\usepackage{jcappub} % for details on the use of the package, please
\usepackage{comment}
\usepackage{color}
\usepackage[normalem]{ulem}

\usepackage{mathrsfs}

\graphicspath{figures}

\renewcommand{\vec}{\mathbf}

\newcommand{\be}{\begin{equation}}
\newcommand{\ee}{\end{equation}}
\newcommand{\bea}{\begin{eqnarray}}
\newcommand{\eea}{\end{eqnarray}}
\newcommand{\bdm}{\begin{displaymath}}
\newcommand{\edm}{\end{displaymath}}
\newcommand{\nn}{\nonumber}

\newcommand{\xv}{\mathbf{x}}

\newcommand{\kv}{\mathbf{k}}

\newcommand{\qv}{\mathbf{q}}
\newcommand{\pv}{\mathbf{p}}

\newcommand{\del}{\delta}

\newcommand{\kmax}{k_{\rm max}}

\newcommand{\Ce}{{\widetilde{\mathbb C}}}

\newcommand{\ie}{{\em i.e.}~}
\newcommand{\eg}{{\em e.g.}~}
\newcommand{\Ms}{\, h^{-1} \, \mathrm{M}_\odot}
\newcommand{\Mpc}{\, h^{-1} \, {\rm Mpc}}

\newcommand{\icMpc}{\, h^{3} \, {\rm Mpc}^{-3}}

\newcommand{\cGpc}{\, h^{-3} \, {\rm Gpc}^3}
\newcommand{\kMpc}{\, h \, {\rm Mpc}^{-1}}

\newcommand{\eq}[1]{eq.~(\ref{#1})}

\def\pin{{\sc{Pinocchio}}}
\def\power{{\sc{PowerI4}}}

\title{Toward a robust inference method for the galaxy bispectrum: likelihood function and model selection}

\author[a,b,c,1]{Andrea Oddo,\note{Corresponding author.}}
\author[b,c,d]{Emiliano Sefusatti,}
\author[e]{Cristiano Porciani,}
\author[b,c,f]{Pierluigi Monaco,}
\author[g]{Ariel G. S\'anchez}

\affiliation[a]{SISSA - International  School  for  Advanced  Studies,  Via  Bonomea  265,  34136 Trieste,  Italy}
\affiliation[b]{Institute for Fundamental Physics of the Universe, Via Beirut 2, 34151 Trieste, Italy}
\affiliation[c]{Istituto Nazionale di Fisica Nucleare, Sezione di Trieste,  via  Valerio  2,  34127 Trieste,  Italy}
\affiliation[d]{Istituto Nazionale di Astrofisica, Osservatorio Astronomico di Trieste, via Tiepolo 11, 34143 Trieste, Italy}
\affiliation[e]{Argelander Institut f\"ur Astronomie der Universit\"at Bonn, Auf dem H\"ugel 71, 53121 Bonn, Germany}
\affiliation[f]{Dipartimento di Fisica, Sezione di Astronomia, Universit\`a di Trieste, via Tiepolo 11, 34143 Trieste, Italy}
\affiliation[g]{Max-Planck-Institut f\"ür extraterrestrische Physik, Postfach 1312, Giessenbachstr, 85741 Garching, Germany}

\emailAdd{andrea.oddo@sissa.it}
\emailAdd{emiliano.sefusatti@inaf.it}
\emailAdd{porciani@astro.uni-bonn.de}

\abstract{ 
The forthcoming generation of galaxy redshift surveys will sample the large-scale structure of the Universe over unprecedented volumes with high-density tracers. This advancement will make robust measurements of three-point clustering statistics possible. In preparation for this improvement, we investigate how several methodological choices can influence inferences based on the bispectrum about galaxy bias and shot noise. We first measure the real-space bispectrum of dark-matter haloes extracted from 298 N-body simulations covering a volume of approximately $1000 \cGpc$. We then fit a series of theoretical models based on tree-level perturbation theory to the numerical data. To achieve this, we estimate the covariance matrix of the measurement errors by using 10,000 mock catalogues generated with the {\pin} code.  We study how the model constraints are influenced by the binning strategy for the bispectrum configurations and by the form of the likelihood function. We also use Bayesian model-selection techniques to single out the optimal theoretical description of our data. We find that a three-parameter bias model combined with Poissonian shot noise is necessary to model the halo bispectrum up to scales of $k_\mathrm{max}\lesssim 0.08 \kMpc$, although fitting formulae that relate the bias parameters can be helpful to reduce the freedom of the model without compromising accuracy. Our data clearly disfavour local Eulerian and local Lagrangian bias models and do not require corrections to Poissonian shot noise. We anticipate that model-selection diagnostics will be particularly useful to extend the analysis to smaller scales as, in this case, the number of model parameters will grow significantly large.
}


\begin{document}

\maketitle

%%%%%%%%%%%%%%%%%%%%%%%%%%%%%%%%%%%%%%%%%%%%%%%%%%%%%%%%%%%%%%%%
%%%%%%%%%%%%%%%%%%%%%%%%%%%%%%%%%%%%%%%%%%%%%%%%%%%%%%%%%%%%%%%%
%%%%%%%%%%%%%%%%%%%%%%%%%%%%%%%%%%%%%%%%%%%%%%%%%%%%%%%%%%%%%%%%
\section{Introduction}

The next batch of galaxy redshift surveys aimed at uncovering the origin of the accelerated expansion of the Universe is designed to map the distribution of high-density tracers of the large-scale structure over comoving volumes of considerable size, \eg \cite{DawsonEtal2016, LaureijsEtal2011, LeviEtal2013}. 
In order to make full use of such observational efforts, it is necessary to go beyond the traditional studies of two-point statistics such as the power spectrum and two-point correlation function. The galaxy bispectrum is the lowest order Fourier-space correlation function quantifying the non-Gaussianity of the galaxy distribution and, as such, has received considerable attention in the last years, with many works aiming at a proper assessment of the additional cosmological information it can provide \cite{SongTaruyaOka2015, TellariniEtal2016, YamauchiYokoyamaTakahashi2017, ByunEtal2017, ChanBlot2017, GagraniSamushia2017, YamauchiYokoyamaTashiro2017, GualdiEtal2018, KaragiannisEtal2018, YankelevichPorciani2019}. In addition, the galaxy bispectrum has been measured most recently in the BOSS galaxy survey and analysed in \cite{GilMarinEtal2015, GilMarinEtal2015B, GilMarinEtal2017} with a first detection of baryonic acoustic oscillations presented in \cite{PearsonSamushia2017} (see also \cite{SlepianEtal2017} for an even earlier detection in the 3-point correlation function).   

An inherent difficulty in the analysis of the galaxy bispectrum derives from the fact that the signal is distributed over a large number of triangle configurations, typically compressed into $\sim 1000$ bins for the range of scales covering the linear and quasi-linear regimes.  Making unbiased inferences thus requires the robust estimation of a high-dimensional covariance matrix (for the bispectrum measurements) and of its inverse (the precison matrix). These are often obtained numerically, using a large set of mock catalogs or N-body simulations \cite{Scoccimarro2000B, SefusattiScoccimarro2005, SefusattiEtal2006, ChanBlot2017, GilMarinEtal2015, GilMarinEtal2017}. In this case, it is necessary to account for statistical and systematic errors in the precision matrix when writing a likelihood function for the model parameters \cite{GilMarinEtal2017}. 

Little work has been done on understanding how several methodological decisions impact the  inference of model parameters based on measurements of the galaxy bispectrum. Even assessing the goodness of fit of bispectrum models is impaired by the limited accuracy with which measurement errors and covariances are known \cite{SefusattiCrocceDesjacques2012}. This unsatisfactory situation provides the first motivation for our study.

In this paper, we investigate the consequences of several assumptions that are routinely made or overlooked in the construction of a likelihood function for fitting a model to measurements of the galaxy bispectrum. In particular, we measure the bispectrum of dark-matter halos extracted from nearly 300 cosmological N-body simulations covering a total volume of approximately $1000 \cGpc$.
Covariance matrices are estimated using an even larger set of 10,000 mock halo catalogs generated with the {\pin} code \cite{MonacoEtal2002}. As we focus on rather large scales, we use a simple perturbative expression at tree level as a benchmark model for the halo bispectrum. In this case, sampling the posterior distribution of the model parameters by means of Monte Carlo Markov Chain (MCMC) simulations is particularly efficient and runs can be performed in a relatively short time. 

We compare the results obtained from adopting different likelihood functions that account for errors in the precision matrix \cite{HartlapEtal2009, SellentinHeavens2016}. Furthermore, we study how the best-fit parameters are affected by the binning strategy adopted for the bispectrum measurements and on how we account for such binning in the theoretical model. 

Finally, we apply model-selection techniques to identify
the optimal number of model parameters that are needed to describe the numerical data. In this case, we focus on simple extensions to our benchmark model given by possible corrections to the shot-noise contributions, and on the possibility of reducing the parameter space by adopting fitting functions or theoretically-motivated relations among the bias parameters.

This work is the first of a series aimed at investigating the joint analysis of the galaxy power spectrum and bispectrum in spectroscopic galaxy surveys. The tests presented here should be regarded as basic but essential steps towards a robust investigation of the full information content of large-scale-structure observables.

This paper is organised as follows. We introduce all numerical data in section~\ref{sec:data} and the adopted likelihood functions in section~\ref{sec:likelihood}. The statistical tools we employ are described in section~\ref{sec:statistics}.  Our results are discussed in section~\ref{sec:results}. Finally, we present our conclusions in section~\ref{sec:conclusions}.

%%%%%%%%%%%%%%%%%%%%%%%%%%%%%%%%%%%%%%%%%%%%%%%%%%%%%%%%%%%%%%%%
%%%%%%%%%%%%%%%%%%%%%%%%%%%%%%%%%%%%%%%%%%%%%%%%%%%%%%%%%%%%%%%%
%%%%%%%%%%%%%%%%%%%%%%%%%%%%%%%%%%%%%%%%%%%%%%%%%%%%%%%%%%%%%%%%
\section{Data}
\label{sec:data}

%%%%%%%%%%%%%%%%%%%%%%%%%%%%%%%%%%%%%%%%%%%%%%%%%%%%%%%%%%%%%%%%
%%%%%%%%%%%%%%%%%%%%%%%%%%%%%%%%%%%%%%%%%%%%%%%%%%%%%%%%%%%%%%%%
\subsection{N-body simulations}

We base our work on the Minerva set of 298 N-Body simulations, first presented in \cite{GriebEtal2016}. Each run evolves the positions and velocities of $1000^3$ dark-matter particles in a periodic cubic box of side $L=1500 \Mpc$ using the \texttt{GADGET-2} code\footnote{http://www.gadgetcode.org/} \citep{Springel2005}.
The flat $\Lambda$CDM background cosmology is determined by the dimensionless Hubble parameter $h=0.695$, the total matter density $\Omega_m=0.285$, and the baryon density $\Omega_b =0.046$. This choice corresponds to the best-fit of the combined analysis of the WMAP results and BOSS DR9 results presented in Table I of  \cite{SanchezEtal2013}. The particle mass in the simulations is thus $m_p \simeq 2.67 \times 10^{11}\Ms$. 
Initial density and velocity perturbations are generated at redshift $z_\mathrm{in}=63$ by displacing the simulation particles from a regular grid using second-order Lagrangian perturbation theory (LPT). The transfer function for the Gaussian linear fluctuations in the matter density is computed with the \texttt{CAMB} code \citep{LewisChallinorLasenby2000} assuming a primordial scalar spectral index of $n_s = 0.9632$ and a r.m.s. matter density fluctuation averaged over spheres of radius equal to 8$\Mpc$  (linearly extrapolated to $z=0$) of $\sigma_8=0.828$.

Dark-matter halos are identified using a standard friends-of-friends algorithm with a linking length of 0.2 times the mean one-dimensional interparticle separation. Unbound particles are removed using the \texttt{SUBFIND} code \cite{SpringelEtal2001}. We only consider halos that contain at least 42 particles, corresponding to a minumum mass of $M \simeq 1.12 \times 10^{13} \Ms$. For simplicity, we limit our study to a single output at redshift $z=1$, as this value is of particular relevance for upcoming spectroscopic galaxy surveys such as Euclid \cite{LaureijsEtal2011} or DESI \cite{AghamousaEtal2016}. The mean number density for the resulting halo population is of $2.13\times 10^{-4}\icMpc$.

%%%%%%%%%%%%%%%%%%%%%%%%%%%%%%%%%%%%%%%%%%%%%%%%%%%%%%%%%%%%%%%%
%%%%%%%%%%%%%%%%%%%%%%%%%%%%%%%%%%%%%%%%%%%%%%%%%%%%%%%%%%%%%%%%
\subsection{Measurements}
\label{ssec:measurements}

We measure the matter and halo power spectra and bispectra from estimates of the Fourier-space overdensity $\del_\kv$ on a grid of linear size 256, obtained with the {\power} code\footnote{\hyperlink{https://github.com/sefusatti/PowerI4}{https://github.com/sefusatti/PowerI4}} described in \cite{SefusattiEtal2016}. The power-spectrum estimator is 
\be
\label{eq:PowerEstimator}
    \hat{P}(k) \equiv \frac{k_f^3}{N_P(k)}\sum_{\qv\in k}\,|\del_{\qv}|^2\,,
\ee
where $k_f\equiv2\pi/L$ is the fundamental wavenumber of the simulation box\footnote{We adopt the following convention for the discrete Fourier transform
\be
    \delta_\kv  \equiv  \int_V \frac{d^3x}{(2\pi)^3}\, e^{-i\kv\cdot\xv}\,\delta(\xv)\,,
\ee
with the inverse given by the series
\be
    \delta(\xv)  \equiv  k_f^3\, \sum_{\kv}\,e^{i\kv\cdot\xv}\,\delta_\kv\,.
\ee
}. The sum runs over all discrete wavevectors $\qv$ in a $k$-bin of size $\Delta k$, \ie with $k-\Delta k/2\le|\qv|< k+\Delta k/2$, and $N_P(k)$ represents their total number. The bispectrum estimator is defined as 
\be
\label{eq:BispEstimator}
    \hat{B}(k_1, k_2, k_3) \equiv \frac{k_f^3}{N_B(k_1, k_2, k_3)}\sum_{\qv_1\in k_1}\sum_{\qv_2\in k_2}\sum_{\qv_3\in k_3} \del_K(\vec q_{123}) \del_{\qv_1}\del_{\qv_2}\del_{\qv_3}\,,
\ee
where $\del_K(\kv)$ denotes the Kronecker delta function (which is 1 when $\kv=0$ and 0 otherwise) and $\qv_{i_1\dots i_n}\equiv\qv_{i_1}+\dots +\qv_{i_n}$. The normalisation factor
\be
\label{eq:BispEstimatorNorm}
    N_B(k_1, k_2, k_3) \equiv \sum_{\qv_1\in k_1}\sum_{\qv_2\in k_2}\sum_{\qv_3\in k_3} \del_K(\vec q_{123}) \,,
\ee 
denotes the total number of wavenumber triplets $(\qv_1,\qv_2,\qv_3)$ forming closed triangles that lie in the ``triangle bin'' defined by the 3-tuple $(k_1,k_2,k_3)$, with the $k_i$'s being the bin centers, and where each bin has a width $\Delta k$. 
In the rest of the paper we will refer to the triplets $(\qv_1,\qv_2,\qv_3)$ formed by wavevectors on the original density grid as ``fundamental triangles'' to distinguish them from the ``triangle bin'' $(k_1,k_2,k_3)$.
An efficient implementation of the algorithm for the bispectrum estimator is described in \cite{Scoccimarro2015}.

We use the same values of $\Delta k$ for both the power spectrum and the bispectrum. In particular, we consider three binning strategies, each defined in terms of three quantities: the bin size $s$ and the center of the first bin $c$ (both given in units of the fundamental wavenumber $k_f$) as well as the total number of bins $N_k$. The bin centers are thus
\be
    k_i=\left[c+(i-1)\,s\right]\,k_f\,,\qquad i=1,\dots, N_k\,.
    \label{eq:binning}
\ee
The adopted values for $s$ and $c$ are shown in table~\ref{tab:binning}.

\begin{table}[t]
 \centering
\bgroup
\def\arraystretch{1}
 \begin{tabular}{l|c|c|c|c}
 \hline\hline
$\Delta k $    & $s$ & $c$ & $N_k$  & $N_{\rm t}$  \\
 \hline
$  k_f$ & 1 & 2.0 & 28   & 2682 \\
$ 2\,k_f$ & 2 & 2.5 & 14  & 399 \\
$ 3\,k_f$ & 3 & 3.0 & 9 & 131 \\
\hline\hline
 \end{tabular}
\egroup
 \caption{Characteristics of the three binning schemes in terms of the bin size ($s$) and the center of the first bin ($c$) -- both given in units of the fundamental wavenumber -- (see the explanation before equation~(\ref{eq:binning})), the total number of bins and the total number of triangle configurations for all bispectrum measurements up to the maximum wavenumber of $k\simeq 0.12\kMpc$.}
 \label{tab:binning}
\end{table}

% -------------------------------

Given a set of bins $\{k_i\}$, we measure the bispectrum for all triangle configurations of sides $k_1\ge k_2\ge k_3$. The total number of triangle configurations for a given value of $\kmax=[c+(N_k-0.5) s]\,k_f$ is given by
\be
    \label{eq:NumberOfTriangles}
    N_{\rm t}=\sum_{i_1=1}^{N_k}\,\sum_{i_2=\lceil (i_1 - c/s)/2\rceil}^{i_1}\,\sum_{i_3=\max[1,\lceil i_1-i_2-c/s \rceil] }^{i_2}\,1;
\ee
This expression includes triangle bins for which the bin {\em centers} do {\em not} satisfy the triangle condition $k_1\le k_2+k_3$ but that still contain closed fundamental triangles.
\footnote{The binning schemes we introduced are defined in such a way that, when Fourier modes are binned in one-dimensional bins based on their modulus -- as is the case for the power spectrum -- there are values of $\kmax$ for which all the modes with $k < \kmax$ are accounted for, regardless of the scheme. For the bispectrum, where triplets of Fourier modes have to be binned, this is exactly true only considering all triangle bins as defined by \eq{eq:NumberOfTriangles}. For instance, the fundamental triangle with sides $(\sqrt{34}, \sqrt{12}, \sqrt{6}) k_f$, contributes to the triangle bins with centers given by $(6,3,2)k_f$ for $s=1$, $(6.5,2.5,2.5)k_f$ for $s=2$, and $(6,3,3)k_f$ for $s=3$. If we limit ourselves to the triangle bins with centers forming a closed triangle, this specific fundamental triangle would only be accounted for in the binning scheme with $s=3$. We will denote ``open bins'' those whose centers do not satisfy the triangle conditions (\eg $(6,3,2)k_f$ for $s=1$ and $(6.5,2.5,2.5)k_f$ for $s=2$) and ``closed bins'' all others (\eg $(6,3,3)k_f$ for $s=3$).}
For the value of $\kmax=0.125\kMpc$ we obtain $N_{\rm t}=2682$, 399 and 131 respectively for $s=1$, 2 and 3. Our measurements in the smallest bin case of $s=1$ are limited to this value of $\kmax$ in order to avoid a rapidly increasing number of configurations.

% -------------------------------

Given the estimator in \eq{eq:BispEstimator} and introducing the notation $t_i=\{k_{i_1},k_{i_2},k_{i_3}\}$ for a generic triangle configuration, the bispectrum covariance can be written in general terms as
\bea
    \label{eq:Bcov}
    C_{ij} 
    & \simeq & \delta^K_{i_1j_1}\,\delta^K_{i_2j_2}\,\delta^K_{i_3j_3}\,\frac{1}{N_B(t_i)}P_{\rm tot}(k_{i_1})\,P_{\rm tot}(k_{i_2})\,P_{\rm tot}(k_{i_3})+5~{\rm perm.}\nn\\
    & & +\del^K_{i_1j_1}\frac{k_f^6}{N_B(t_i)N_B(t_j)}\sum_{\qv_1\in k_{i_1}}\sum_{\qv_2\in k_{i_2}}\sum_{\qv_3\in k_{i_3}}\sum_{\pv_2\in k_{j_2}}\sum_{\pv_3\in k_{j_3}}\del_K(\qv_{123})\del_K(\qv_1\!\!-\!\!\pv_{23})\nn\\
    & & \times B_{\rm tot}(\qv_1,\qv_2,\qv_3)B_{\rm tot}(\qv_1,\pv_2,\pv_3)+2~{\rm perm.}\nn\\
    & & +\del^K_{i_1j_1}\frac{k_f^6}{N_B(t_i)N_B(t_j)}\sum_{\qv_1\in k_{i_1}}\sum_{\qv_2\in k_{i_2}}\sum_{\qv_3\in k_{i_3}}\sum_{\pv_2\in k_{j_2}}\sum_{\pv_3\in k_{j_3}}\del_K(\qv_{123})\del_K(\qv_1\!\!-\!\!\pv_{23})\nn\\
    & & \times P_{\rm tot}(\qv_1)T_{\rm tot}(\qv_2,\qv_3,\pv_2,\pv_3)+2~{\rm perm.}\nn\\
    & & +\frac{k_f^9}{N_B(t_i)N_B(t_j)}\sum_{\qv_1\in k_{i_1}}\sum_{\qv_2\in k_{i_2}}\sum_{\qv_3\in k_{i_3}}\sum_{\pv_1\in k_{j_1}}\sum_{\pv_2\in k_{j_2}}\sum_{\pv_3\in k_{j_3}}\del_K(\qv_{123})\del_K(\pv_{123})\nn\\
    & & \times T_{6,{\rm tot}}(\qv_1,\qv_2,\qv_3,\pv_1,\pv_2,\pv_3)\,,
\eea
where we assumed the thin-shell approximation ($k_i\gg \Delta k$) to be valid for the Gaussian contribution on the r.h.s. and $P_{\rm tot}$, $B_{\rm tot}$, $T_{\rm tot}$ and $T_{6,{\rm tot}}$ represent, respectively, the power spectrum, bispectrum, trispectrum and the connected 6-point function including shot-noise. In particular, for a purely Poisson shot-noise we have
\be
    \label{eq:Ptot}
    P_{\rm tot}(k)=P(k)+\frac1{(2\pi)^3\bar{n}}\,,
\ee
with $\bar{n}$ being the halo number density, and 
\be
    \label{eq:Btot}
    B_{\rm tot}(k_1,k_2,k_3)=B(k_1,k_2,k_3)+\frac1{(2\pi)^3\bar{n}}\left[P(k_1)+P(k_2)+P(k_3)\right]+\frac1{(2\pi)^6\bar{n}^2}\,.
\ee

\begin{figure}[t]
    \centering
    \includegraphics[width=0.95\textwidth]{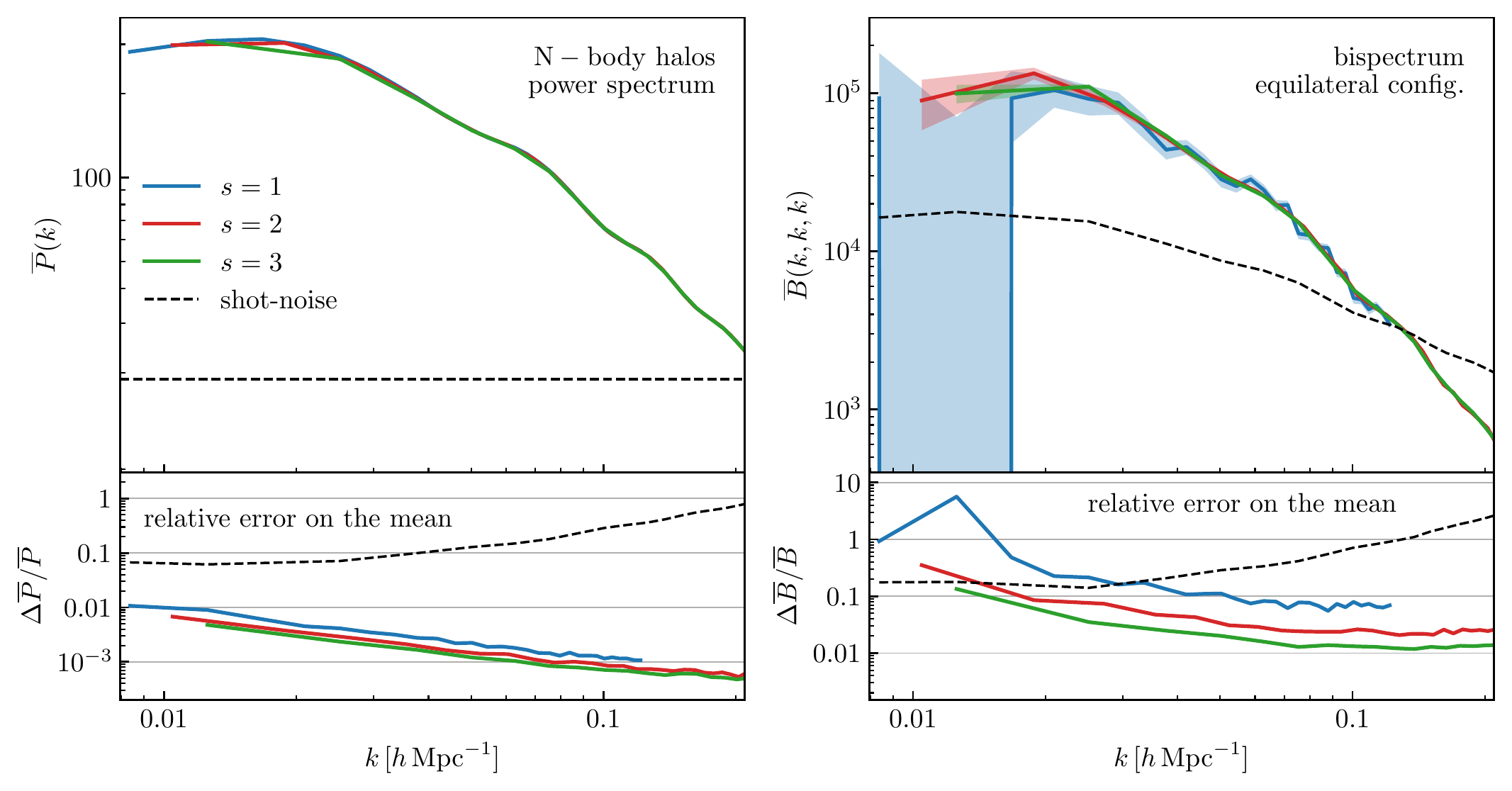}
    \caption{Average measurements of the halo power spectrum (left panels) and of the equilateral configurations of the bispectrum (right panels) over the 298 N-body simulations for the three binning schemes, along with the Poisson shot-noise contribution (dashed lines, already subtracted from the measurements). Shaded areas denote the error on the mean. The bottom panels show the relative error on the mean (solid lines) and the relative shot-noise contribution (dashed lines). Note that the measurements of the power spectrum and of the equilateral bispectrum for the different binning schemes are consistent with each other, with uncertainties being sub-percent in the case of the power spectrum. Also evident is the relatively larger shot-noise contribution to the bispectrum measurements w.r.t. the power spectrum ones.}
    \label{fig:measurements_ps_bseq}
\end{figure}

Figure~\ref{fig:measurements_ps_bseq} shows the mean halo power spectrum and the mean equilateral configurations of the bispectrum from the 298 N-body simulations for the three binning choices. The bottom panels show the relative error on the mean. 
Due to the large number of simulations we consider, the average of our measurements reaches subpercent precision in the determination of the halo power spectrum even on the relatively large scales we are interested in.
Notice that this is of the same order of possible systematic effects introduced by the set-up and implementation of the N-body solver (see \eg \cite{SchneiderEtal2016} for a quantification of systematic errors on the matter power spectrum). In the case of the halo bispectrum, the relative error on the mean ranges between one and several tens of percent, depending on the binning. 
Also shown in figure~\ref{fig:measurements_ps_bseq} is the Poisson shot-noise contribution expected for the two statistics. 
For $P(k)$, the constant shot-noise contribution is generally subdominant w.r.t. the signal, but becomes of comparable size for $k\sim 0.2\kMpc$. For the equilateral bispectrum configurations the (scale-dependent) shot-noise contribution is relatively larger, accounting for at least 10\% of the signal at large scales and becoming dominant already around $k\sim 0.12\kMpc$. 

\begin{figure}[p!]
    \centering
    \includegraphics[width=0.9\textwidth]{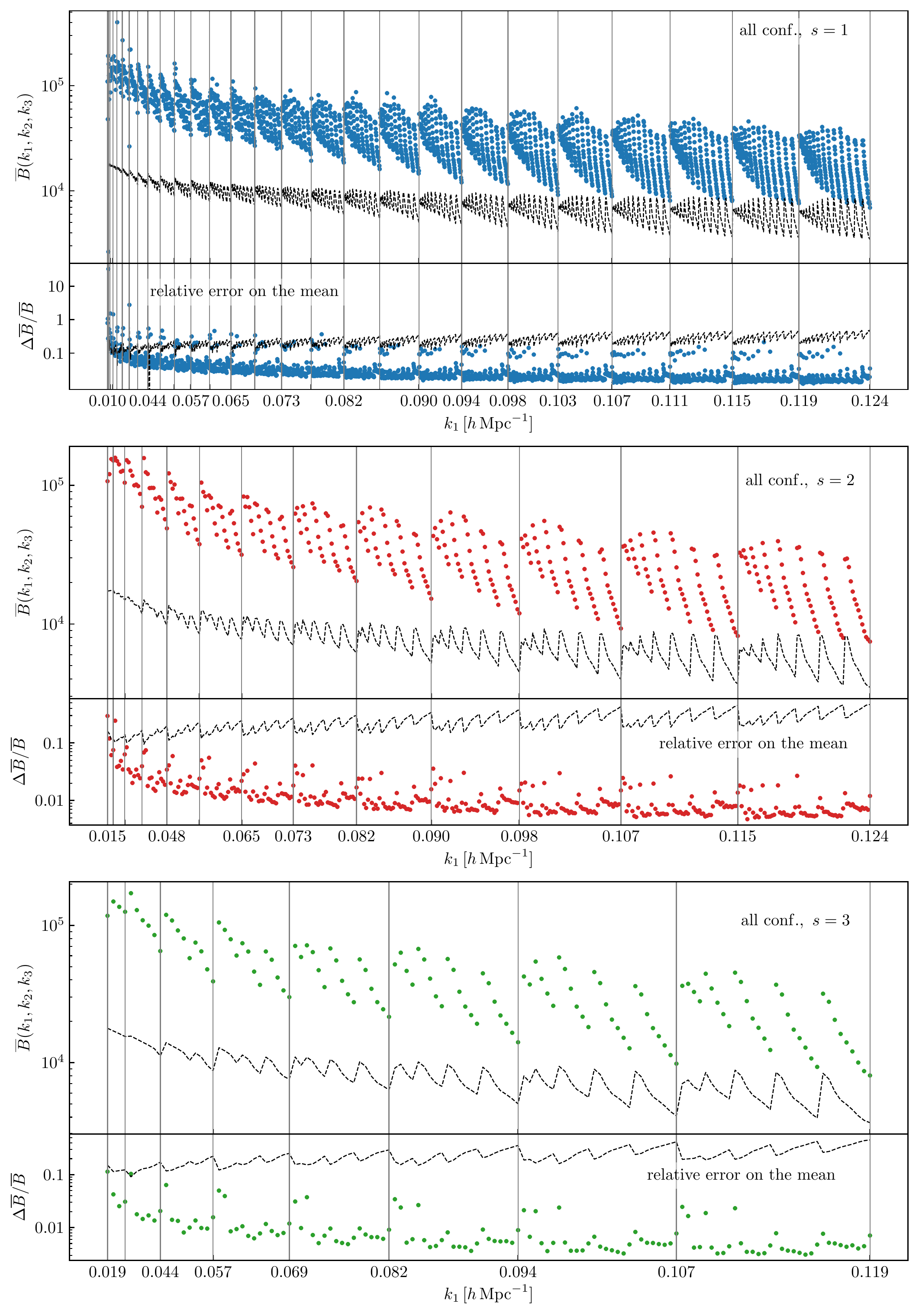}
    \caption{Measurements of the halo bispectrum from the N-body simulations for the three binning schemes. In each panel, the bottom plot shows the relative error on the mean. Dotted black lines show the absolute (top) and relative (bottom) Poisson shot-noise contribution to the measurements.}
    \label{fig:measurements_bs_all}
\end{figure}

Similarly, figure~\ref{fig:measurements_bs_all} shows the mean of {\em all} measured triangle configurations and the relative error on the mean for the three binning schemes. The order of the triangles appearing in these plots (and in several others in the rest of the paper) matches the sums in \eq{eq:NumberOfTriangles} and corresponds to increasing the value of the sides $\{k_1,k_2,k_3\}$ with the constraint $k_1\ge k_2\ge k_3$. The ticks on the horizontal axis and the corresponding vertical lines mark equilateral configurations where the value of $k_1$ changes. It follows that, in between two ticks, all points correspond to triangles with the same $k_1$, while $k_2$ and $k_3$ assume all allowed values.
Again, the Poisson shot-noise contribution is shown with a dashed line in the upper half of each panel while its relative size appears in the bottom half.

%%%%%%%%%%%%%%%%%%%%%%%%%%%%%%%%%%%%%%%%%%%%%%%%%%%%%%%%%%%%%%%%
%%%%%%%%%%%%%%%%%%%%%%%%%%%%%%%%%%%%%%%%%%%%%%%%%%%%%%%%%%%%%%%%
\subsection{Mock halo catalogs}
\label{ssec:mocks}

In addition to the full N-body simulations, we make use of a larger set of $10,000$ mock halo catalogs generated with the {\pin} code
\cite{MonacoTheunsTaffoni2002, MonacoEtal2013, MunariEtal2017}
using the same cosmological model and simulation settings. {\pin} 
uses third-order Lagrangian Perturbation Theory to shift matter particles
%relies on particle displacements obtained with third-order Lagrangian Perturbation Theory and 
and relies on a set of criteria, based on the ellipsoidal-collapse model, to group them into halos. Note that 298 {\pin} realisations have been obtained using the same random seeds for the initial conditions of the Minerva simulations, thereby allowing a one-to-one comparison not affected by sample variance. 
Both the N-body simulations and the {\pin} mocks have been introduced and used in a series of papers \cite{LippichEtal2019, BlotEtal2019, ColavincenzoEtal2019} aimed at comparing several methods for the production of approximate catalogs in terms of their predictions for the 2-point correlation function, power spectrum, and bispectrum, along with their respective covariance properties. We refer the reader to these works for a first assessment of the accuracy of the {\pin} mocks and to \cite{Monaco2016} for a general review of approximate methods.

The main results of references \cite{LippichEtal2019, BlotEtal2019, ColavincenzoEtal2019} are derived defining the mock halo catalogs in terms of a mass threshold that provides the same halo number density, on average, as the reference N-body simulations. We will make here a different choice and set the mass threshold by matching the clustering amplitude of the halos in real space as determined by the {\em total} halo power spectrum,  including shot noise, \eq{eq:Ptot}, since this is the quantity determining the bispectrum Gaussian variance. The mass threshold, in this case, controls the overall amplitude via both the linear bias and the number density.  
This matching is crucial to minimize any systematic difference between the covariance matrices for the bispectrum (and the power spectrum) extracted from the {\pin} and the N-body simulations. In fact, as shown in \eq{eq:Bcov}, the Gaussian contribution to the error on the bispectrum measurements, representing the leading term at large scales, depends essentially on $P_{\rm tot}(k)$.  

\begin{figure}[t!]
    \centering
    \includegraphics[width=0.95\textwidth]{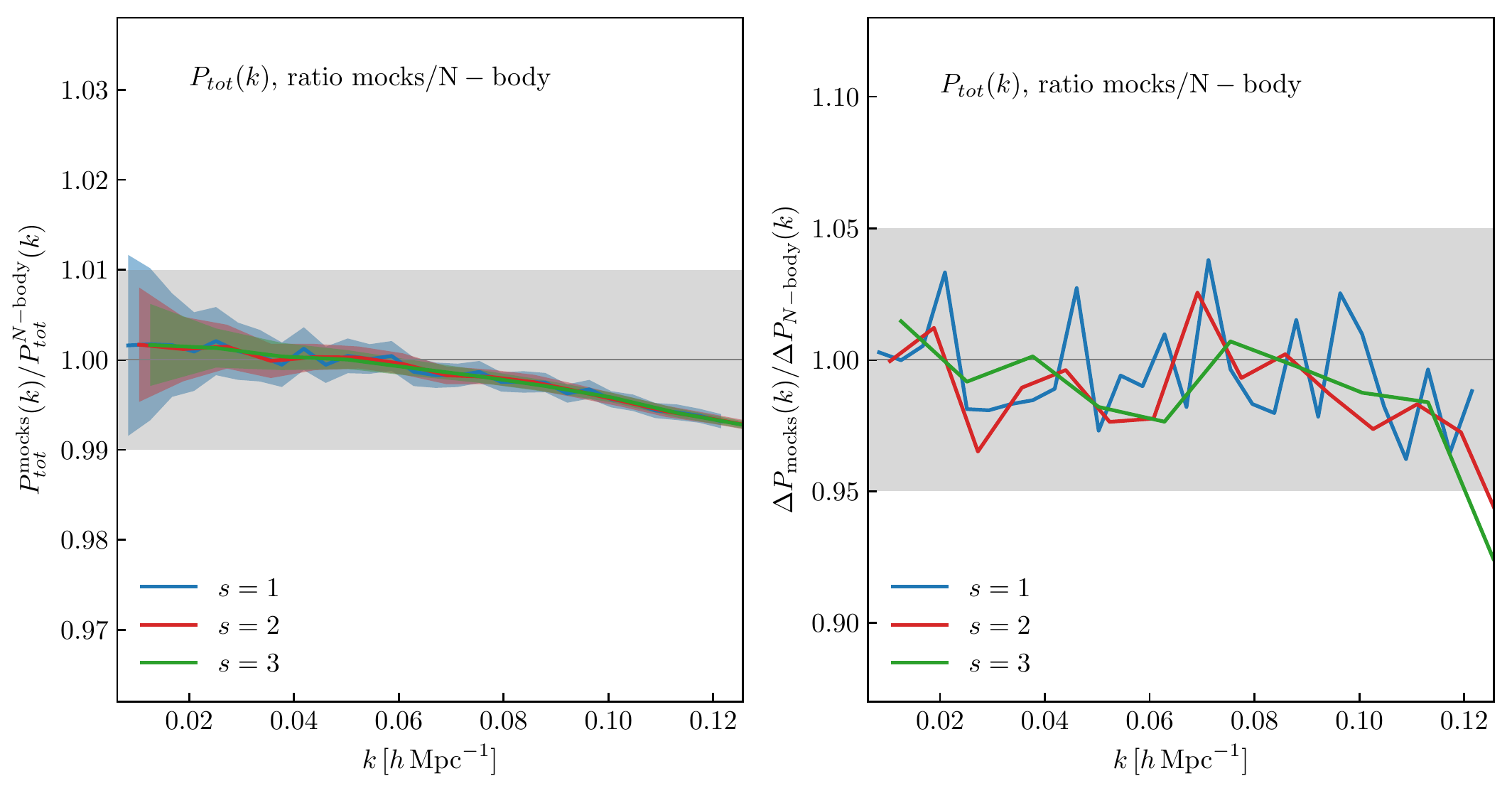}
    \caption{Left: ratio of the mean total halo power spectrum estimated from the {\pin} mocks to the same quantity measured from the N-body simulations for the 298 realisations sharing the same initial conditions. Different colors represent the different binning schemes in $k$, with the shaded regions denoting the corresponding error on the mean. The total power spectra are consistent with each other at the percent level at least up to $0.12\,h\, {\rm Mpc}^{-1}$.
    Right: ratio between the standard deviations estimated from the {\pin} mocks  and from the N-body simulations, again limited to the 298 realisations sharing the initial seeds. The standard deviations are consistent at the $\sim 5$ percent level up to $0.12\,h\, {\rm Mpc}^{-1}$. }
    \label{fig:PIN_vs_MIN_power}
\end{figure}

The solid lines in the left-hand-side panel of figure~\ref{fig:PIN_vs_MIN_power} show the ratio between the mean $P_{\rm tot}(k)$ estimated from the {\pin} mocks and from the N-body simulations. This comparison is limited to the 298 realisations sharing the same initial conditions. 
Different binning schemes are represented with different colors and the shaded regions denote the corresponding error on the mean of the mocks measurements.
The total power in the {\pin} mocks matches the result from the N-body simulations to better than one percent up to at least $k\sim 0.12 \kMpc$. These are the scales we are interested in. Notice that the {\em shot-noise subtracted} power spectra show, instead, a discrepancy of about 2-3\% at the largest scales, compensated by a similar discrepancy in the shot-noise contribution. The ratio between the power-spectrum variance estimated from the {\pin} mocks and from the N-body simulations is shown in the right-hand-side panel of figure~\ref{fig:PIN_vs_MIN_power}, again only using the 298 realisations with the same initial seeds. The ratio scatters around one, with a few percent deviations.

\begin{figure}[t!]
    \centering
    \includegraphics[width=0.95\textwidth]{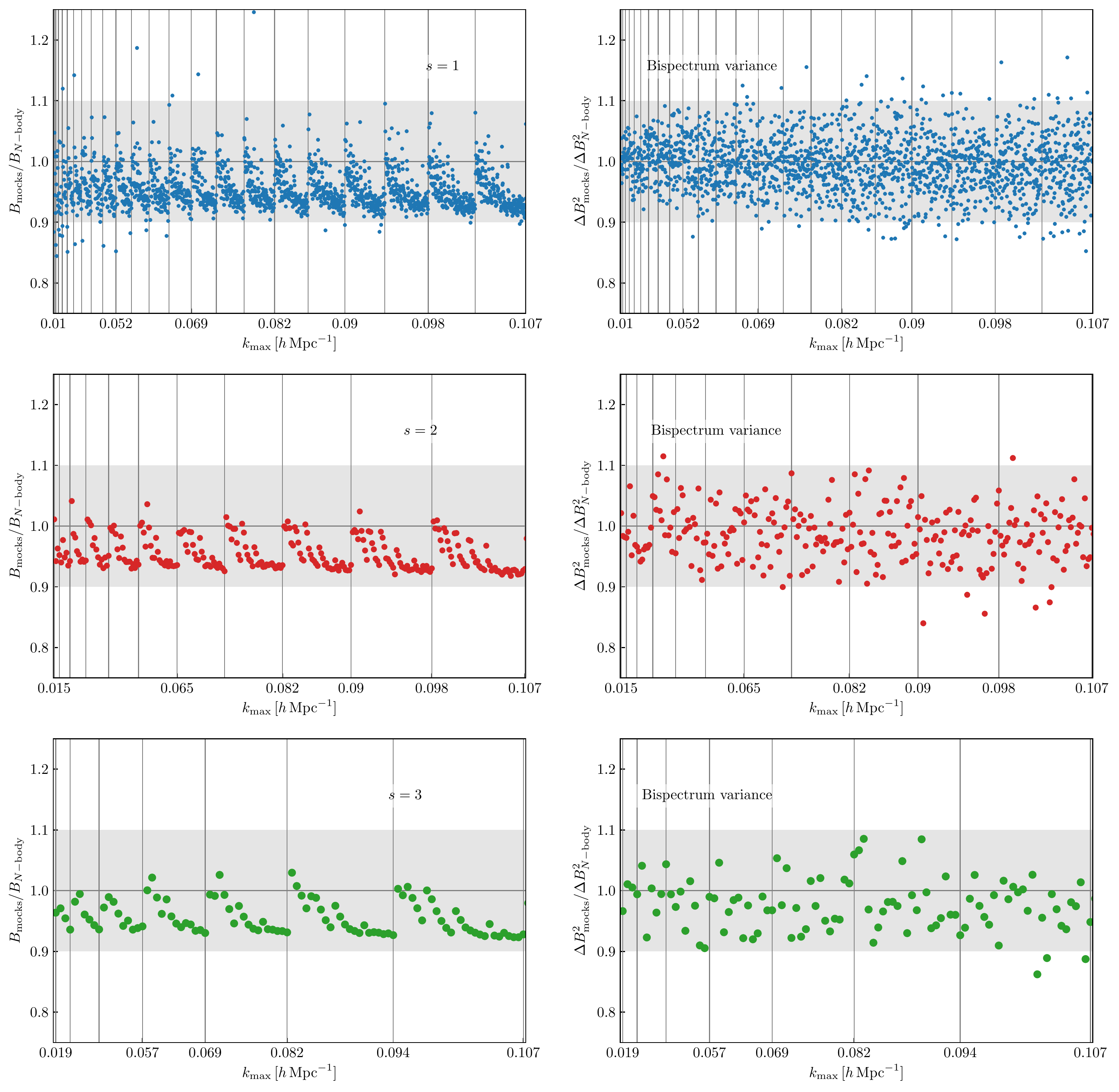}
    \caption{The left-hand-side panels display the ratio of the mean bispectrum measured from the {\pin} mocks (with Poisson shot-noise subtracted) to the corresponding quantity measured from the N-body simulations for the three binning schemes adopted in this work ($s=1$, 2 and 3 from top to bottom). The right-hand side panels show the ratio of the bispectrum variance measured in the mocks to the one measured from the simulations. Both comparisons consider only the 298 realisations with matched initial conditions.}
    \label{fig:PIN_vs_MIN_bisp}
\end{figure}

Figure~\ref{fig:PIN_vs_MIN_bisp} shows a similar comparison but in terms of the bispectrum and its variance. The left-hand-side panels show the ratio between the mean bispectra measured from the {\pin} mocks and from the N-body simulations for the three binning schemes. In this case, we consider measurements corrected for Poisson shot-noise, as in \eq{eq:Btot}. Notice that the bispectrum from the mocks is suppressed with respect to the one from the simulations by about 6-7\% with some dependence on the triangle shape that follows from the discrepancy in the large-scale power spectra mentioned above, in addition to the small-scale suppression due to LPT displacements. The right-hand-side panels show that the variance of the bispectrum in the {\pin} mocks reproduces that in the N-body simulations with a scatter of about 10\% but with no significant systematic error, except for a slight suppression at the few percent level visible in the $s=3$ measurements. 

%%%%%%%%%%%%%%%%%%%%%%%%%%%%%%%%%%%%%%%%%%%%%%%%%%%%%%%%%%%%%%%%
%%%%%%%%%%%%%%%%%%%%%%%%%%%%%%%%%%%%%%%%%%%%%%%%%%%%%%%%%%%%%%%%
\subsection{Covariance}

\begin{figure}[p]
    \centering
    \includegraphics[width=0.85\textwidth]{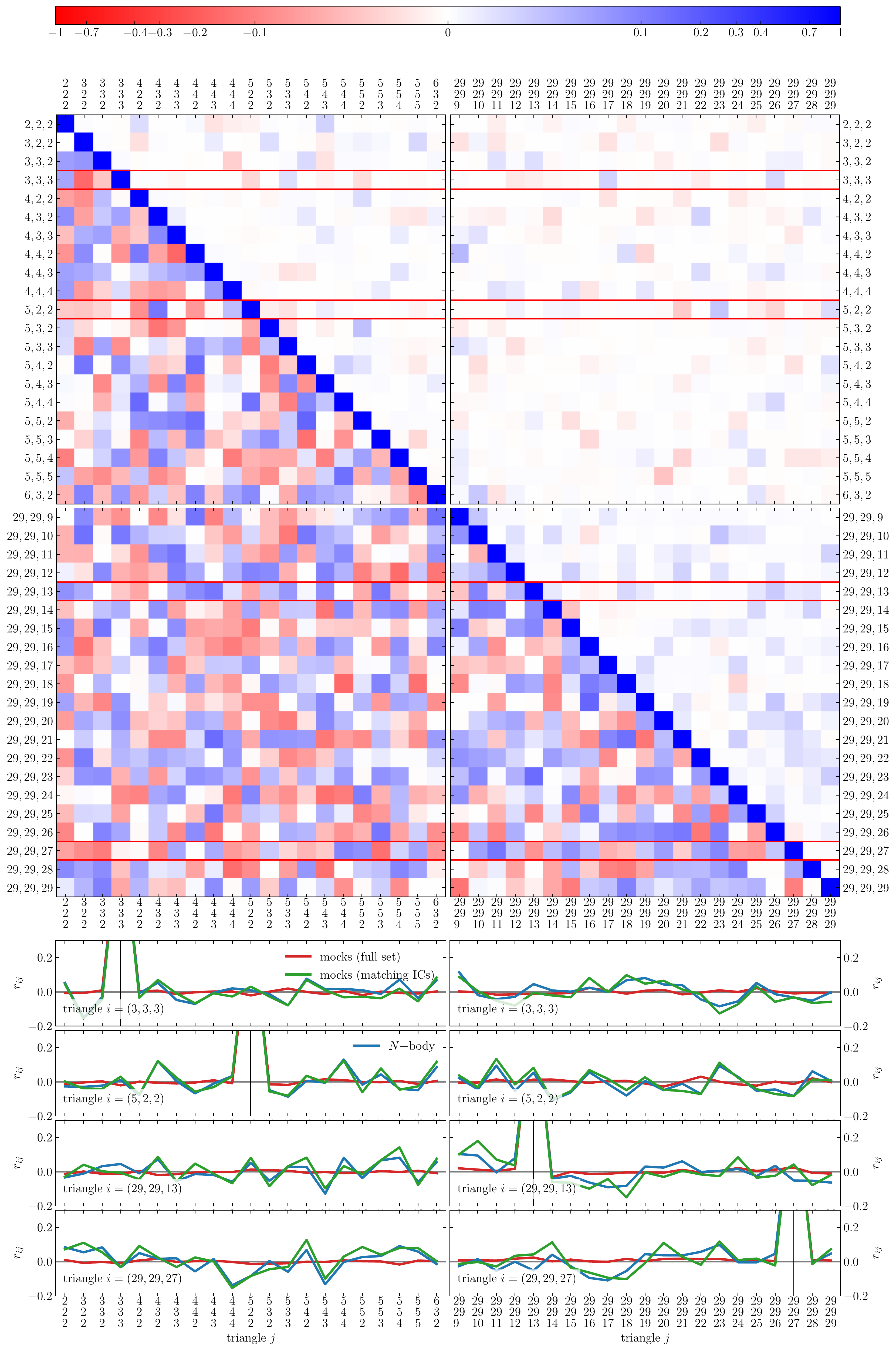}
    \caption{The cross-correlation coefficients $r_{ij}$, \eq{eq:rij}, of the bispectrum covariance matrix estimated from the 10,000 {\pin} mocks compared with those estimated from the 298 N-body simulations for the binning with $s=1$. The top four panels represent the four corners of the matrix as defined by the ordering described above in \eq{eq:NumberOfTriangles}. Each row and column corresponds to a triangle whose sides are given in units of the fundamental wavenumber as a triplet of integers. The smallest-scale triangle in this case is $\{29,29,29\}k_f=\{0.12,0.12,0.12\}\kMpc$. Matrix elements above the diagonal are estimated from the 10,000 mocks, while those below  are estimated from the 298 simulations. Bottom panels show the rows marked above with red borders, comparing the estimates from 298 simulations (blue), from the 298 mocks with matching seeds (green) and from the full set of 10,000 mocks (red).}
    \label{fig:covariance_s1}
\end{figure}

Figure~\ref{fig:covariance_s1} shows the cross-correlation matrix
\be
\label{eq:rij}
    r_{ij}\equiv \frac{C_{ij}}{\sqrt{C_{ii}\,C_{jj}}}\, ,
\ee
of the bispectrum measurements extracted from the {\pin} and the N-body simulations using bins with $s=1$. In particular, the top four panels represent the four corners of the $r_{ij}$ matrix as defined by the ordering described above in \eq{eq:NumberOfTriangles}, illustrating the correlation properties of large-scale and small-scale triangles. Each row and column corresponds to a triangle configuration whose sides are given in units of the fundamental wavenumber as a triplet of integers. The smallest-scale triangle in this case is $\{29,29,29\}k_f=\{0.12,0.12,0.12\}\kMpc$. Matrix elements above the diagonal are estimated from the 10,000 mocks, while those below are estimated from the 298 simulations. Clearly, both covariances appear to be dominated by diagonal terms but show very different noise levels for the off-diagonal elements.
The bottom panels compare some rows of the matrix $r_{ij}$ (that have been highlighted in the top plots with red borders) obtained from different datasets. They allow a more direct and quantitative comparison of the results from the 10,000 {\pin} mocks (red lines), the 298 N-body simulations (blue lines), and the corresponding 298 mocks with matched initial conditions (green lines). Notice that the noise in the $r_{ij}$ elements extracted from this last dataset reproduces quite closely the noise coming from simulations. On the other hand, the off-diagonal cross-correlations from the full set of mocks are very close to zero, with differences well below the 5\% level. 

\begin{figure}[p]
    \centering
    \includegraphics[width=0.85\textwidth]{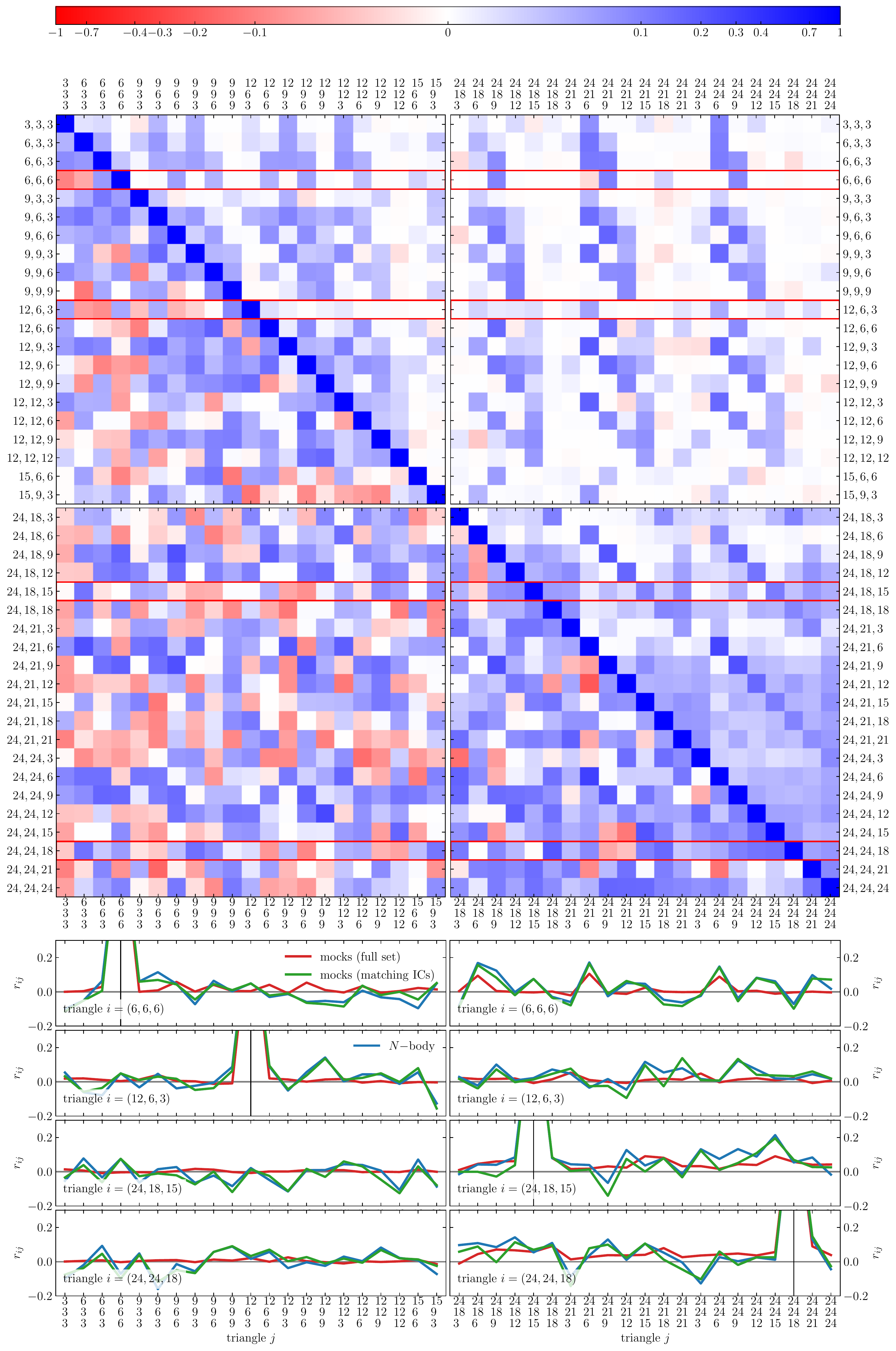}
    \caption{As in figure~\ref{fig:covariance_s1} but for $s=3$. Off-diagonal correlations become more evident with a binning scheme with larger $s$.}
    \label{fig:covariance_s3}
\end{figure}

It would be misleading, based on this simple inspection, to conclude that off-diagonal contributions to the  covariance matrix of the bispectrum can be safely ignored.
A counterexample is shown in figure~\ref{fig:covariance_s3} where we consider bins with $s=3$.
Here the top panels show again the four corners of the covariance matrix up to the same maximum wavenumber, with the last triangles being $\{8,8,8\}\Delta k=\{24,24,24\}k_f=\{0.1,0.1,0.1\}\kMpc$. Several off-diagonal elements of the matrix $r_{ij}$ assume values of 10-20\%. They correspond to different triangles that share one or more sides and it is expected that the bispectrum covariance receives contributions from the non-Gaussian terms in this case.
These features are present as well in the covariance estimated from the 298 simulations, but the noise affecting them is also of the order of 10-20\%. 
Correlations among different triangles are present for any binning scheme and in the same overall amount. 

In order to quantify the importance of off-diagonal terms, let us consider the cumulative signal-to-noise ratio for the bispectrum defined as
\be
    \label{eq:StoN}
    \left(\frac{S}{N}\right)^2=\sum^{N_{\rm t}(\kmax)}_{i,j}B_i\,C_{ij}^{-1}\,B_j\,,
\ee
where the indices $i$ and $j$ run over all triangle configurations having no sides larger than $\kmax$.

\begin{figure}[t!]
    \centering
    \includegraphics[width=0.95\textwidth]{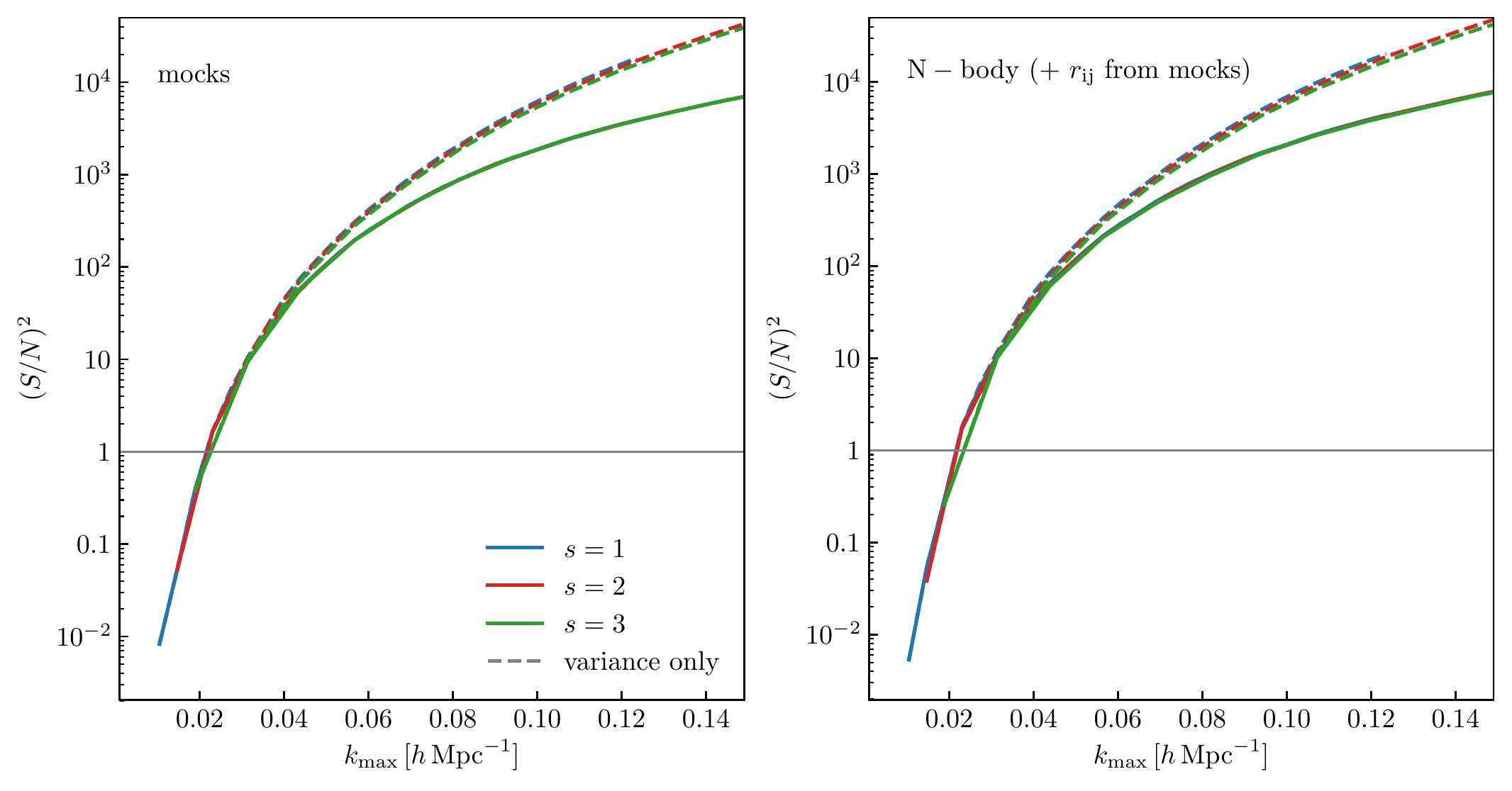}
    \caption{Cumulative signal-to-noise ratio for the bispectrum measured from the {\pin} mocks (left panel) and from the N-body simulations (right panel). In the latter case, the full covariance matrix is estimated from the variance measured from the N-body runs and the cross-correlation coefficients $r_{ij}$ estimated again from the {\pin} mocks, since a direct measurement of the covariance matrix only from the N-body simulations would be too noisy. The solid lines are obtained considering the full covariance matrix (for the simulations we use the cross-correlation coefficients estimated from the 10,000 mocks). For comparison, we also show results obtained using only the diagonal part of the covariance matrix (dashed lines). 
    Results for the three binning schemes we adopt are shown with different colors. As expected, they almost perfectly coincide.}
    \label{fig:StoN}
\end{figure}

It is interesting to see how this quantity changes if one neglects the contribution from off-diagonal terms in the covariance.
This test is performed in the left-hand-side panel of figure~\ref{fig:StoN} where we use the ``signal'' and the ``noise'' extracted from the 10,000 mocks. In this case we can expect a sufficiently precise determination of the covariance matrix (and its inverse), with a residual statistical error on the order of a few percent.
Here the solid and dashed lines represent, respectively, the results using the full covariance and the diagonal part alone. Note that off-diagonal terms become increasingly more important as $\kmax$ grows, causing a reduction of $(S/N)^2$ by a factor of two at $\kmax\sim 0.08 \kMpc$.
In the right-hand-side panel, instead, we repeat the test by using the bispectrum measurements $B_i$ and their variance $\Delta B_i$ from the N-body simulations. In this case, the covariance matrix is obtained using the relation $C_{ij}=\Delta B_i\,\Delta B_j\,r_{ij}$ where the cross-correlation matrix is estimated from the 10,000 mocks. The results are essentially the same as in the left-hand-side panel. 

The comparison between mocks and simulations over the subset of realisations sharing the initial seeds performed in terms of the bispectrum variance, figure~\ref{fig:PIN_vs_MIN_bisp}, as well as in terms of cross-correlation coefficients $r_{ij}$, constitutes our main justification for using the covariance from the larger set of 10,000 {\pin} runs in the analysis of the simulation measurements. In what follows, we will assume that any systematic error in the determination of the bispectrum covariance matrix based on the approximate halo mocks is negligible, but we will comment on those few instances where a  small residual systematic error could affect our results.

%%%%%%%%%%%%%%%%%%%%%%%%%%%%%%%%%%%%%%%%%%%%%%%%%%%%%%%%%%%%%%%%
%%%%%%%%%%%%%%%%%%%%%%%%%%%%%%%%%%%%%%%%%%%%%%%%%%%%%%%%%%%%%%%%
%%%%%%%%%%%%%%%%%%%%%%%%%%%%%%%%%%%%%%%%%%%%%%%%%%%%%%%%%%%%%%%%
\section{Model inference}
\label{sec:likelihood}

In this section we describe the likelihood function adopted for our analysis, along with the theoretical model and the treatment of possible, residual statistical errors in the estimation of the covariance matrix. 

\subsection{Theoretical model}
\label{sec:theomod}
We use a theoretical model for the halo bispectrum that only includes leading-order contributions in standard perturbation theory (SPT, see \cite{BernardeauEtal2002} for a review).
This choice is motivated by the fact that we consider only large-scale measurements, with wavevectors $k<\kmax=0.12 \kMpc$ at redshift $z=1$. 
Specifically, we consider the expression
\bea
    B_h(\kv_1, \kv_2, \kv_3) & =  & b_1^3\, B_\mathrm{TL}(\kv_1, \kv_2, \kv_3) + b_2 b_1^2\, \Sigma(\kv_1, \kv_2, \kv_3) + 2\gamma_2 b_1^2\, K(\kv_1,\kv_2,\kv_3)+ \nn\\
    & & + \frac{1}{(2 \pi)^3\bar{n}} (1+\alpha_1)\, b_1^2\,\left[P_\mathrm{L}(k_1)+P_\mathrm{L}(k_2)+P_\mathrm{L}(k_3)\right] +  \frac{1}{(2 \pi)^6\bar{n}^2} (1+\alpha_2)\;,
    \label{eq:bispmodel}
\eea
that derives from assuming a second-order bias expansion \cite{FryGaztanaga1993, ChanScoccimarroSheth2012, BaldaufEtal2012} with parameters $b_1$ (linear bias), $b_2$ (quadratic bias) and $\gamma_2$ (tidal bias, for which we use the notation in \cite{ChanScoccimarroSheth2012, EggemeierScoccimarroSmith2018}) as well as possible corrections to Poissonian shot noise \cite{SeljakHamausDesjacques2009, HamausEtal2010, PaechEtal2017, GinzburgDesjacquesChan2017} quantified by the coefficients $\alpha_1$ and $\alpha_2$.
Consistent with the leading-order approximation, we ignore any scale-dependence of these shot-noise/stochastic parameters \cite{GinzburgDesjacquesChan2017, EggemeierScoccimarroSmith2018}.
In eq.~(\ref{eq:bispmodel}),
\be
    B_\mathrm{TL}(\kv_1, \kv_2,\kv_3) = 2 F_2(\kv_1, \kv_2) P_\mathrm{L}(k_1) P_\mathrm{L}(k_2) + 2~\mathrm{perm.}\,,
\ee
denotes the tree-level SPT solution for the matter bispectrum, where 
\be
    F_2(\kv_1, \kv_2)=\frac57+\frac12\frac{\kv_1 \cdot \kv_2}{k_1\,k_2}\left(\frac{k_1}{k_2}+\frac{k_2}{k_1}\right)+\frac27\left(\frac{\kv_1 \cdot \kv_2}{k_1\,k_2}\right)^2
\ee 
is the usual second-order kernel and $P_\mathrm{L}(k)$ is the linear power spectrum for the perturbations in the matter density. Similarly,
\be
    \Sigma(\kv_1,\kv_2,\kv_3) = P_\mathrm{L}(k_1)P_\mathrm{L}(k_2) + 2~\mathrm{perm.}\;,
\ee
and 
\be
    K(\kv_1, \kv_2, \kv_3) = \left[\left(\hat {\kv}_1 \cdot \hat {\kv}_2\right)^2 - 1\right]P_\mathrm{L}(k_1)P_\mathrm{L}(k_2) + 2~\mathrm{perm.}\;.
\ee
Finally, it is worth noticing that, in alternative perturbative approaches \cite{Senatore2015}, additional terms appear in the tree-level bispectrum of biased tracers \cite{AnguloEtal2015, FujitaEtal2016A, NadlerPerkoSenatore2018}. However, as they present a $k^2$ scaling similar to 1-loop corrections in SPT we neglect them in this work.  

We fix the cosmological parameters to the actual values of the N-body simulations and only fit the five bias plus shot-noise parameters. This allows us to pre-compute the functions $P_\mathrm{L}, B_\mathrm{TL}, \Sigma$ and $K$, as well as the fully averaged sum of the three linear power spectra in eq.~(\ref{eq:bispmodel}), and thus minimize the time required to evaluate the model and the likelihood function. Note that, since we fit the model to measurements of the bispectrum from individual simulations, we use the halo number density $\bar{n}$ measured in each box to evaluate the Poissonian shot-noise contribution.

Although eq.~(\ref{eq:bispmodel}) contains five tunable parameters, it is important to assess how many of them can be justified by the data as a function of $\kmax$. For this reason, we consider different models obtained by reducing the parameter space in eq.~(\ref{eq:bispmodel}) as follows (see also table~\ref{tab:models} for a compact summary):
\begin{itemize}
    \item ${\mathcal M}_1$: The most basic model assumes a linear-bias relation ($b_2=\gamma_2=0$) and Poissonian shot noise ($\alpha_1=\alpha_2=0$); 
    
    \item ${\mathcal M}_{1f}$: This is still a 1-parameter model where both quadratic parameters $b_2$ and $\gamma_2$,\footnote{\label{fn:b2}The definition of the tidal-bias operator adopted in \cite{LazeyrasEtal2016} differs from ours. They use the ``square'' of the traceless tidal field to define it while we use the second-order Galileon operator. As a result, our second-order bias, $b_2$, relates to theirs, $\tilde{b}_2$, as $b_2=\tilde{b}_2+\frac{4}{3}\gamma_2$.} are expressed as functions of the linear bias parameter, respectively through the fitting function $\tilde b_2(b_1) = 0.412 - 2.143\, b_1 + 0.929\, b_1^2 + 0.008\, b_1^3$ provided by \cite{LazeyrasEtal2016} (see also \cite{HoffmannBelGaztanaga2015, HoffmannBelGaztanaga2017}) and assuming local Lagrangian biasing\footnote{ Several recent studies provide evidence of small systematic deviations from this relation, with $\gamma_2$ being slightly more negative but still linearly related to $b_1$  \cite{SaitoEtal2014, LazeyrasSchmidt2018, AbidiBaldauf2018}. Larger deviations have been measured by    \cite{ModiCastorinaSeljak2017}} \cite{CatelanPorcianiKamionkowski2000}, \ie $\gamma_2=-\frac{2}{7}(b_1-1)$ \cite{ChanScoccimarroSheth2012, BaldaufEtal2012}. For the shot-noise corrections we assume the Poisson prediction $\alpha_1=\alpha_2=0$;
    
    \item ${\mathcal M}_{2b_2}$: 2-parameter model with $b_1$ and $b_2$ free to vary while $\gamma_2=-\frac{2}{7}(b_1-1)$ and  $\alpha_1=\alpha_2=0$;
    
    \item ${\mathcal M}_{2\gamma_2}$: 2-parameter model with $b_1$ and $\gamma_2$ free to vary while $\tilde b_2(b_1)$ follows the fit in \cite{LazeyrasEtal2016} and  $\alpha_1=\alpha_2=0$;  
    
    \item ${\mathcal M}_{2\mathrm{loc}}$: 2-parameter, local model with $b_1$ and $b_2$ free to vary while $\gamma_2=0$ and  $\alpha_1=\alpha_2=0$.
    
    \item ${\mathcal M}_3$: 3-parameter model with all bias parameters free to vary while $\alpha_1=\alpha_2=0$; this is used as reference model;
    
    \item ${\mathcal M}_4$: 4-parameter model with additional freedom in the description of shot noise; a single shot-noise correction parameter $\alpha_1=\alpha_2$ is allowed to vary, as assumed for instance in \cite{GilMarinEtal2015, GilMarinEtal2017}, in addition to the three bias parameters;
    
    \item ${\mathcal M}_5$: 5-parameter model where all parameters in eq.~(\ref{eq:bispmodel}) are free to vary.
    
\end{itemize}

It is worth stressing that our main goal here is to assess the constraining power of the halo bispectrum as a function of $\kmax$ without considering other data that set additional constraints and break degeneracies among the model parameters. We will discuss the combination with the halo power spectrum in our future work.
Note that forthcoming galaxy redshift surveys will span much smaller volumes than our N-body simulations but also deal with a substantially higher number-density of tracers ($\gtrsim 10^{-3} h^{3}$ Mpc$^{-3}$). 
A redshift bin of size $\Delta z=0.2$, that is large enough to properly measure BAO features along the line of sight, corresponds to a volume at most of about 10-12$\cGpc$ both for DESI~\cite{AghamousaEtal2016} or Euclid~\cite{EuclidCollaboration2019}. 
Therefore, our bispectrum measurements are subject to a smaller sample variance and larger shot-noise corrections with respect to what will be available from future galaxy samples.

\begin{table}[!htbp]
\centering
\begin{tabular}{l||c|c|c|c|c}
\hline
\hline
Model                    & $b_1$        & $b_2$        & $\gamma_2$      & $\alpha_1$   & $\alpha_2$   \\
\hline
$\mathcal M_1$           & $\checkmark$ & 0            & 0               & 0            & 0            \\
$\mathcal M_{1f}$        & $\checkmark$ & $\tilde b_2(b_1) + \frac{4}{3}\gamma_2(b_1)$   & $-\frac{2}{7}(b_1 - 1)$ & 0            & 0            \\
$\mathcal M_{2\rm loc}$  & $\checkmark$ & $\checkmark$ & 0               & 0            & 0            \\
$\mathcal M_{2b_2}$      & $\checkmark$ & $\checkmark$ & $-\frac{2}{7}(b_1 - 1)$ & 0            & 0            \\
$\mathcal M_{2\gamma_2}$ & $\checkmark$ & $\tilde b_2(b_1) + \frac{4}{3}\gamma_2$   & $\checkmark$    & 0            & 0            \\
$\mathcal M_3$           & $\checkmark$ & $\checkmark$ & $\checkmark$    & 0            & 0            \\
$\mathcal M_4$           & $\checkmark$ & $\checkmark$ & $\checkmark$    & $\checkmark$ & $\alpha_1$   \\
$\mathcal M_5$           & $\checkmark$ & $\checkmark$ & $\checkmark$    & $\checkmark$ & $\checkmark$ \\
\hline
\hline
\end{tabular}
\caption{Summary of the models analysed in this paper. A checkmark $\checkmark$ highlights the parameters that are left free to vary. The remaining ones are set to the value indicated  in the table. Here, $\tilde{b}_2(b_1)$ is the fitting formula by \cite{LazeyrasEtal2016} for the alternative quadratic bias coefficient introduced in footnote \ref{fn:b2}.}
\label{tab:models}
\end{table}
%%%%%%%%%%%%%%%%%%%%%%%%%%%%%%%%%%%%%%%%%%%%%%%%%%%%%%%%%%%%%%%%
%%%%%%%%%%%%%%%%%%%%%%%%%%%%%%%%%%%%%%%%%%%%%%%%%%%%%%%%%%%%%%%%
\subsection{Binning effects}
\label{ssec:binningtheory}

In practice, the bispectrum is estimated by averaging over a number of triangle configurations (see section~\ref{ssec:measurements}). It is therefore imperative that the theoretical predictions are treated in the same way.
Since our measurements are performed in periodic boxes, the average over the bin is in fact a sum over fundamental triangles:
\be
    \label{eq:binnedbisp}
    B_\mathrm{bin}(k_1, k_2, k_3) =\frac{1}{N_B(k_1, k_2, k_3)}\sum _{\vec q_1 \in k_1} \sum _{\vec q_2 \in k_2} \sum _{\vec q_3 \in k_3}\delta_K(\vec q_{123}) B_h(\vec q_1, \vec q_2, \vec q_3)\;.
\ee
For the tree-level model introduced in eq.~(\ref{eq:bispmodel}), this operation can be performed exactly in a relatively short time (see also \cite{YankelevichPorciani2019}). 
However, in general, this is a non-trivial problem. If the prediction for a single configuration includes loop corrections, it becomes prohibitively expensive to evaluate the average (although, in certain perturbative schemes, it should be possible to approximately replace the sum with an integral).
We explore here the impact of a pragmatic solution that minimizes the number of evaluations of the function $B_h$. The idea is to replace the bin average with the bispectrum evaluated at a single ``effective'' triangle configuration defined in terms of the triangle bin  $\{k_1$, $k_2$, $k_3\}$ and the bin size $\Delta k$.
Generalizing \cite{SefusattiCrocceDesjacques2010} to discrete Fourier transforms, a simple choice would be
\be
    \label{eq:effectiveks1}
    \widetilde{k}_i(k_1;k_2,k_3) = \frac{1}{N_B(k_1, k_2, k_3)}\sum _{\vec q_1 \in k_1} \sum _{\vec q_2 \in k_2} \sum _{\vec q_3 \in k_3} q_i\, \delta_K(\vec q_{123})\;.
\ee
Notice that, with this definition, an ``equilateral bin'' with $k_1=k_2=k_3$ would correspond to an equilateral ``effective'' configuration with $\widetilde{k}_1=\widetilde{k}_2=\widetilde{k}_3$. However, most of the fundamental triangles $\{\qv_1, \qv_2, \qv_3\}$ contributing to such a bin are not equilateral. 
We take this into account by introducing a second effective configuration in which we sort the sidelengths of the fundamental triangles before averaging them (see also \cite{YankelevichPorciani2019}). For instance, the largest effective side would be given by
\be
    \label{eq:effectiveks2}
    \widetilde{k}_l(k_1,k_2,k_3) = \frac{1}{N_B(k_1, k_2, k_3)}\sum _{\vec q_1 \in k_1} \sum _{\vec q_2 \in k_2} \sum _{\vec q_3 \in k_3} \max(q_1,q_2,q_3)\, \delta_K(\vec q_{123})\,,
\ee
and similar definitions can be written for the middle and smallest values $\widetilde{k}_m$ and $\widetilde{k}_s$. 

\begin{figure}
    \centering
    \includegraphics[width=0.95\textwidth]{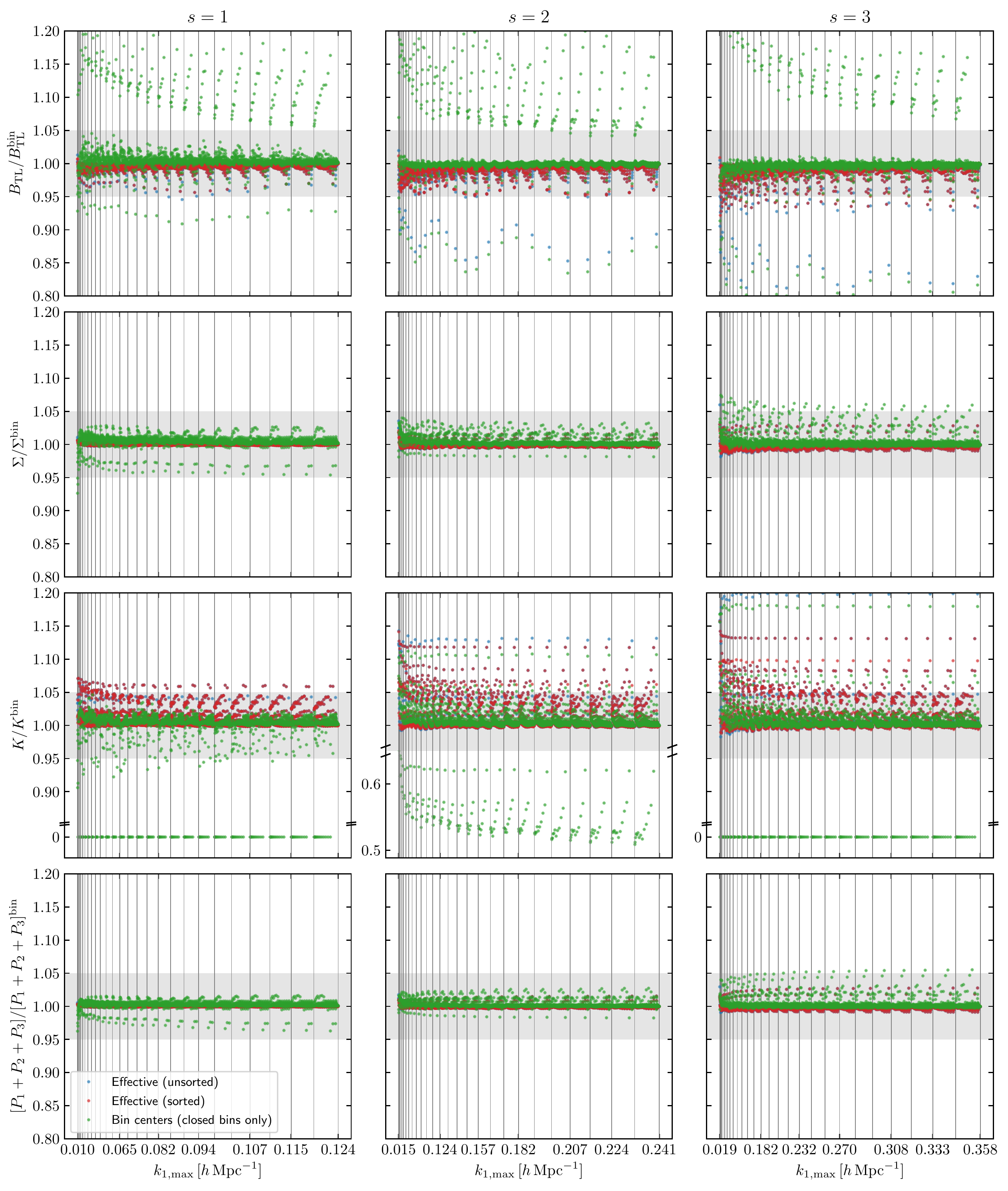}
    \caption{Comparison of the bispectrum model evaluated at particular triangle configurations of wavenumbers and its full average over the triangle bins, equation~(\ref{eq:binnedbisp}). We consider two different definitions of effective triangles obtained by sorting or not the sides of a fundamental triangle before averaging them over the bin, see equations~(\ref{eq:effectiveks1}) and (\ref{eq:effectiveks2}). A third option is obtained by taking the bin center for each leg of the triangular bin, but this procedure is applied only for bins where the bin centers form a closed triangle (``closed bins''). 
    Each column corresponds to a different binning scheme, $s=1,2,3$ from left to right. Each row refers to a different contribution to the tree-level model for the bispectrum given in \eq{eq:bispmodel}.
    Notice the broken $y$-axes in the third row of panels, where the large deviations are due to the fact that the terms $[(\hat {\kv}_1 \cdot \hat {\kv}_2)^2 - 1]$ become exactly zero for collinear triangle bins in binning schemes with $s=1,3$; this does not happen for the scheme with $s=2$ since the bin centers defined there never form a collinear triangle with $k_1 = k_2 + k_3$.
    The effective-sorted method performs generally better, while evaluating the model at the bin centers performs the worst.}
    \label{fig:prediction_comp}
\end{figure}

In figure \ref{fig:prediction_comp}, we show how much the bispectrum evaluated at the effective configurations differs from the correct bin average given in \eq{eq:binnedbisp}. We plot all the different contributions to the tree-level bispectrum separately.
For $s=1$, discrepancies are generally at the few percent level, with a small subset of configurations exceeding 5\% only for the $K$ term. As expected, such differences grow with the bin size but remain smaller than 5\%, now with the exception of $K$ and the matter contribution $B_{\rm TL}$, \ie both shape-dependent terms. For the specific case of squeezed isosceles triangles (\ie those configurations with bin {\em centers} given by $k_1=k_2$ and $k_3=\Delta k$) we notice that using the averages of the sorted wavenumbers as in \eq{eq:effectiveks2} works much better than the other case. For completeness, in figure \ref{fig:prediction_comp}, we also consider theoretical predictions for the bispectrum directly evaluated at the bin centers $\{k_1, k_2, k_3\}$, clearly limited to those triangle bins whose triplet centers form a closed triangle (``closed bins'').
In general, using the bin center performs much worse than any of the two effective solutions we introduced above, and moreover this treatment is not well defined for open bins. For instance, for collinear triangles with $k_1=k_2+k_3$, it always gives $K=0$ for the $s=1$ and $s=3$ cases. 

%%%%%%%%%%%%%%%%%%%%%%%%%%%%%%%%%%%%%%%%%%%%%%%%%%%%%%%%%%%%%%%%
%%%%%%%%%%%%%%%%%%%%%%%%%%%%%%%%%%%%%%%%%%%%%%%%%%%%%%%%%%%%%%%%
\subsection{Likelihood function}
\label{sec:likshape}
The likelihood function $\mathcal{L}$ of a hypothesis given some data is proportional to the probability of obtaining the data under the assumption that the hypothesis is true.
The simplest and most-commonly made assumption is to treat the data as generated by an unbiased estimator that produces Gaussian measurement errors.
For a set of $N_\mathrm{t}$ bispectrum configurations, this gives 
\be
    \ln \mathcal L = -\frac{1}{2}\sum_{i=1}^{N_\mathrm{t}}\sum_{j=1}^{N_\mathrm{t}} \delta B_i\, C_{ij}^{-1} \,\delta B_j\equiv -\frac{1}{2}\, \delta \vec B^T \cdot \mathbb C^{-1} \cdot \delta \vec B=-\frac{\chi^2}{2}\;,
    \label{eq:gausslik}
\ee
with $\delta B_i =  B_i^\mathrm{obs} - B^\mathrm {th}_i$ where $B_i^\mathrm{obs}$ denotes the observed (binned) bispectrum for the triangle configuration $t_i$ and $B^\mathrm{th}_i$ is the corresponding theoretical prediction. 
The second expression on the rhs adopts a compact notation expressed in terms of the bispectrum data vector $\vec B$ and  the corresponding covariance matrix $\mathbb C$. 
In our analysis, however, the covariance matrix  is numerically estimated from the 10,000 mocks described in section~\ref{ssec:mocks} using their sample covariance. Despite the large number of realisations, the resulting estimate, $\tilde{\mathbb{C}}$ is still plagued by statistical errors that generate two effects.
First, on average, they produce a bias in the precision matrix \cite{Anderson2003}. Secondly, they lead to a loss of information in the parameter-inference process (\ie the resulting posterior probability distributions of the model parameters are, on average, broader than in the absence of noise) \cite{DodelsonSchneider2013, TaylorJoachimiKitching2013, TaylorJoachimi2014, PercivalEtal2014, SellentinHeavens2017}.
It has been shown  that considering Gaussian errors and marginalizing over the (unknown) population covariance matrix (by assuming an independence Jeffreys prior) given the noisy estimate leads to the non-Gaussian likelihood \cite{SellentinHeavens2016}:
\be
    \ln \mathcal L = -\frac{N_M}{2} \ln{\left[ 1 + \frac{\delta {\vec B} \cdot \Ce^{-1} \cdot \delta {\vec B}}{N_M-1} \right]} + \ln{\left( \frac{\bar c_p}{\sqrt{\det\Ce}} \right)},
    \label{eq:SellentinLikelihood}
\ee
where $N_M$ denotes the number of simulations used to estimate the covariance matrix and
$\bar c_p$ is a normalization constant that does not depend on the model parameters and we can neglect. This can be interpreted as a generalisation of the multivariate $t$-distribution. Note that eq.~(\ref{eq:SellentinLikelihood}) has the same numerical complexity as the multivariate Gaussian in eq.~(\ref{eq:gausslik}) and can thus be easily used to run MCMC simulations. This is our reference method. 
Another approach frequently used in the literature is to adopt a Gaussian likelihood with
an unbiased estimator of the precision matrix
obtained by rescaling the inverse of the sample
covariance matrix \cite{Anderson2003, HartlapSimonSchneider2007}:
\be
    \label{eq:anderson}
    \hat{\mathbb{C}}^{-1} = \frac{N_M-N_{\rm t}-2}{N_M-1}\, \widetilde{\mathbb{C}}^{-1}\;.
\ee
We also consider this possibility.
In our case, $N_M=10,000$ and $N_\mathrm{t}$ varies with the binning strategy. For $s=1$ and $\kmax\simeq 0.12\kMpc$, we have $N_\mathrm{t}\sim 2,700$ and the correction is of the order of 27\% and thus not negligible. For $s=2$ and 3 and the same $\kmax$, instead, it amounts to a few percent or less.
Note that the credibility intervals of the model parameters obtained using either eq.~(\ref{eq:SellentinLikelihood}) or a Gaussian likelihood with the rescaled precision matrix given in eq.~(\ref{eq:anderson}) need to be corrected in order to account for the loss of information due to the uncertainty in the covariance matrix, see \eg \cite{PercivalEtal2014, SellentinHeavens2017}. We will revisit this procedure in Section~\ref{sec:results}.

A related issue is assessing to what degree eq.~(\ref{eq:gausslik}) provides a good approximation to the actual likelihood function.
Departures from the ideal case could originate for different reasons. In the first place, we can expect that bispectrum estimates at the largest scales have a non-Gaussian probability distribution simply because they behave like the third power of the Gaussian field $\del_h(\qv)$ \cite{Scoccimarro2000B, HahnEtal2019}. In fact,
while at small scales these measurements get contributions from a large number of Fourier modes and Gaussianity is recovered due to the central limit theorem, for $k$-bins close to the fundamental wavenumber $k_f$, only a few modes contribute, and a certain degree of non-Gaussianity could be present. In addition, one can expect a non-Gaussian distribution to arise from the small-scale non-linear evolution and non-linear bias. 
However, after inspecting the noisy probability density function of our 298 bispectrum measurements, as well as the values of skewness and kurtosis of the distributions of each individual triangle configuration normalised by their standard errors, we could not detect any significant departure from a Gaussian distribution at the scales that are relevant for this work.

Our ultimate goal is to assess the suitability of different theoretical models for the halo bispectrum on large scales. We thus fit the models to the measurements extracted from our $N_R=298$ N-body simulations. Since the different realisations are statistically independent, we write the total likelihood function as the product of the likelihoods of the individual measurements or, equivalently,
\be
    \ln \mathcal L_\mathrm{\rm tot} = \sum_{\alpha=1}^{N_R} \ln \mathcal L_\alpha\;.
\ee
For the partial likelihood functions, we use eq.~(\ref{eq:SellentinLikelihood}), neglecting the normalisation constant, 
as our benchmark method.
For comparison, we also use eq.~(\ref{eq:gausslik}) combined with eq.~(\ref{eq:anderson}). Note that, in the Gaussian case, this procedure gives the same parameter constraints as fitting the mean of our measurements with a suitably rescaled covariance matrix (\eg \cite{SefusattiCrocceDesjacques2012}).  This does not hold true in general, though.

%%%%%%%%%%%%%%%%%%%%%%%%%%%%%%%%%%%%%%%%%%%%%%%%%%%%%%%%%%%%%%%%
%%%%%%%%%%%%%%%%%%%%%%%%%%%%%%%%%%%%%%%%%%%%%%%%%%%%%%%%%%%%%%%%
\subsection{Prior probabilities and posterior distribution}
\begin{table}[t]
    \centering

    \begin{tabular}{l|c|c}
        \hline
        \hline
        Parameter & Broad prior (uniform) & Narrow prior (uniform) \\
        \hline
        $b_1$ &  $[0.5,5]$ & $[0.5,5]$\\
        $b_2$ &  $[-5,5]$ & $[-5,5]$\\
        $\gamma_2$ & $[-5,5]$ & $[-5,5]$\\
        $\alpha_1$ &  $[-10,10]$ & $[-1,1]$\\
        $\alpha_2$ &  $[-100,100]$ & $[-1,1]$\\
        \hline
        \hline
    \end{tabular}
    
    \caption{Uniform prior intervals of the model parameters. For the shot-noise parameters, two different priors are used: the broader one does not influence the inference process while the narrower one makes sure that shot-noise corrections are positive and deviations from Poisson noise are small.}
    \label{tab:priors}
\end{table}

We assume uniform priors for all the model parameters (see Table~\ref{tab:priors}). 
More specifically, for the shot-noise parameters, we consider two possibilities. We first consider non-informative, very broad priors that do not influence the inference process (so that it is fully determined by the bispectrum data).
In the second set of priors for the shot noise parameters, we force the shot noise to be positive (\ie $\alpha_{1,2}>-1$) and make sure that corrections are not larger than the Poisson terms.

We evaluate the posterior distribution by means of Monte Carlo Markov Chain simulations using the Python affine-invariant sampler \texttt{emcee} \cite{ForemanMackeyEtal2013}. For each run, we use 100 walkers to sample the parameter space.
To make sure that they are sufficiently converged, we stop the simulations after 50 integrated autocorrelation times \cite{GoodmanWeare2010}.

%%%%%%%%%%%%%%%%%%%%%%%%%%%%%%%%%%%%%%%%%%%%%%%%%%%%%%%%%%%%%%%%
%%%%%%%%%%%%%%%%%%%%%%%%%%%%%%%%%%%%%%%%%%%%%%%%%%%%%%%%%%%%%%%%
%%%%%%%%%%%%%%%%%%%%%%%%%%%%%%%%%%%%%%%%%%%%%%%%%%%%%%%%%%%%%%%%
\section{Statistical methods: goodness of fit and model comparison}
\label{sec:statistics}

In this section, we provide a concise introduction to the statistical tools we use to determine the goodness of fit between observed and predicted results as well as to compare the performance of models with a different number of parameters.

%%%%%%%%%%%%%%%%%%%%%%%%%%%%%%%%%%%%%%%%%%%%%%%%%%%%%%%%%%%%%%%%
%%%%%%%%%%%%%%%%%%%%%%%%%%%%%%%%%%%%%%%%%%%%%%%%%%%%%%%%%%%%%%%%
\subsection{Model selection}
\label{sec:MS}

We consider a set of candidate models $\mathcal M_k\in \mathscr{M}$ for the halo bispectrum (each characterized by the parameters set $\boldsymbol{\theta}_k$) and use different criteria to select which of them ``better'' agrees with the measurements from our simulations. 
Two major threads run through the statistical literature on Bayesian model selection that go under the names of ``explanatory'' and ``predictive'' modelling. Let us consider a particular phenomenon or mechanism that gives rise to noisy data (the Data Generating Process or DGP). A model is an often imperfect collection of mathematical and statistical rules giving rise to an output that  resembles the actual data in important ways. Explanatory modelling tests theoretically motivated hypotheses and looks for the ``true model'' (\ie the DGP or approximations thereof) by examining posterior probabilities given the observed data \cite{KassRaftery1995}. Predictive modelling, on the other hand, searches for the model that makes the best prediction of future observations generated by the same DGP as the observed data. 
This is based on the posterior predictive probability distribution function which gives the probability of future measurements conditional on the observed data: 
\be
    p(\mathrm{new\ data}|\mathrm{data},\mathcal M_k)=\int p(\mathrm{new\ data}|\boldsymbol{\theta}_k,\mathcal M_k)\,P(\boldsymbol{\theta}_k|\mathrm{data})\,\mathrm{d}\boldsymbol{\theta}_k\;,
    \label{eq:predictivepdf}
\ee
where $P$ denotes the posterior distribution of $\boldsymbol{\theta}_k$ given the data.
For finite and noisy measurements, the two concepts possibly lead to different conclusions. In this work, we make use of one model selection criterion from each class.

%%%%%%%%%%%%%%%%%%%%%%%%%%%%%%%%%%%%%%%%%%%%%%%%%%%%%%%%%%%%%%%%
\paragraph{Bayes factors from the Savage-Dickey density ratio}

In order to evaluate the evidence that the data provide in favour of $\mathcal M_j$ with respect to $\mathcal M_k$, explanatory modelling generally relies on calculating the Bayes Factor (BF), \ie the ratio between the probabilities of observing the data under the models,
\be
    \mathrm{BF}_{jk}=\frac{p(\mathrm{data}|\mathcal M_j)}{p(\mathrm{data}|\mathcal M_k)}=
    \frac{\int \mathcal{L}(\mathrm{data}|\boldsymbol{\theta}_j)\,\pi(\boldsymbol{\theta}_j|\mathcal M_j)\,\mathrm{d}\boldsymbol{\theta}_j}{\int \mathcal{L}(\mathrm{data}|\boldsymbol{\theta}_k)\,\pi(\boldsymbol{\theta}_k|\mathcal M_k)\,\mathrm{d}\boldsymbol{\theta}_k}\;,
    \label{BFdef}
\ee
where $\pi(\boldsymbol{\theta}_j|\mathcal M_j)$ denotes the prior for the model parameters. 
This quantity coincides with the ratio between the marginal likelihoods, \ie between the a priori predictions that the models make about the probability of observing the data in the experiment.
The posterior odds of model $\mathcal M_j$ relative to model $\mathcal M_k$ are obtained by multiplying the prior odds times the Bayes factor $\mathrm{BF}_{jk}$.
Ref. \cite{Jeffreys1961} recommends that Bayes factors larger than 3, 10, and 100 should be used to speak of substantial, strong and decisive evidence in favour of model $\mathcal M_j$ against model $\mathcal M_k$.
Note that Bayes factors account for model complexity and  automatically penalise more complicated models with respect to simpler ones. 

The main drawback of applying this machinery to our bispectrum study is that Bayes factors are notoriously challenging to compute from MCMC simulations. This is because the MCMC method sparsely samples regions of parameter space where the likelihood function is relatively low that might give important (if not dominant) contributions to the integrals in equation~(\ref{BFdef}) \cite{KassRaftery1995}.
However, the situation improves dramatically if we limit our analysis to properly nested models.
Model $\mathcal M_j$ is said to be properly nested under $\mathcal M_{k}$ if: (i) the parameter set $\boldsymbol{\theta}_j \subset \boldsymbol{\theta}_{k}$ (i.e. $\mathcal M_j$ is obtained by fixing each additional parameter in $\mathcal M_{k}$ to a constant value, i.e. $\boldsymbol{\psi}=\mathbf{c}$ where $\boldsymbol{\psi}$ denotes the set of parameters that are free to vary in $\mathcal{M}_k$ and are fixed in $\mathcal{M}_j$); (ii) the prior distributions in the models satisfy  $\lim_{\boldsymbol{\psi}\to\mathbf{c}} \pi(\boldsymbol{\theta}_{k}|\mathcal M_{k})=\pi(\boldsymbol{\theta}_j|\mathcal M_j)$; and (iii) the likelihood functions in the models satisfy the relation $\mathcal{L}(\mathrm{data}|\boldsymbol{\theta}_j,\mathcal M_j)=\mathcal{L}(\mathrm{data}|\boldsymbol{\theta}_j, \mathbf{c},\mathcal M_{k})$. In this case, the Bayes factor is given by \cite{Dickey1971}
\be
    \mathrm{BF}_{jk}=\frac{\int P(\boldsymbol{\theta}_j,\boldsymbol{\psi}=\mathbf{c}|\mathcal M_{k})\,\mathrm{d}\boldsymbol{\theta}_j}{\int \pi(\boldsymbol{\theta}_j,\boldsymbol{\psi}=\mathbf{c},|\mathcal M_{k})\,\mathrm{d}\boldsymbol{\theta}_j}\;,
\ee
which is known as the Savage-Dickey density ratio and is relatively easy to compute from a MCMC simulation. See \eg \cite{Trotta2007} for an application to cosmology.

%%%%%%%%%%%%%%%%%%%%%%%%%%%%%%%%%%%%%%%%%%%%%%%%%%%%%%%%%%%%%%%%
\paragraph{Deviance Information Criterion}

Given a dataset, a model $\mathcal M_k$, and a particular set of values for the model parameters $\hat{\boldsymbol{\theta}}_k$, we define the deviance statistic as
\be
    D=2 \log C(\mathrm{data}) -2 \log \mathcal{L}(\mathrm{data}| \hat{\boldsymbol{\theta}}_k,\mathcal M_k)\;,
\ee
where the fully specified function $C(\mathrm{data})$ does not depend on the candidate model.
Frequentist model assessment is based on the difference of the log-likelihoods between a model and the saturated model (that perfectly fits all data) and can thus be also formulated in terms of the difference of the deviances (note that, apart from the constant $C(\mathrm{data})$, $D$ coincides with the $\chi^2$ statistic for measurements with Gaussian noise).
Ref. \cite{Dempster1974} suggested to use the posterior distribution of $D$ as a measure of goodness of fit in the Bayesian framework. 
\cite{SpiegelhalterEtal2002} formalised this concept by introducing the Deviance Information Criterion (DIC) as a method for model selection. 
This is based on the following statistic: 
\begin{equation}
    \mathrm{DIC}=\langle D \rangle_\mathrm{post}+p_D\;,
    \label{eq:dicpd}
    \end{equation}
where $\langle D \rangle_\mathrm{post}$ denotes the posterior expectation of the deviance and
$p_D=\langle D \rangle_\mathrm{post} -D(\langle \boldsymbol{\theta}\rangle_\mathrm{post})$ is a Bayesian measure of model complexity that gives an estimate of the effective number of model parameters.  Models associated with a lower DIC are better supported by the data. Starting from the definitions given above, we can write the DIC metric as $\mathrm{DIC}=D(\langle\boldsymbol{\theta}\rangle_\mathrm{post})+2p_D$.  This notation makes it clear that  the DIC is based on a trade-off between model accuracy and complexity.
Fit quality is measured by plugging the posterior mean of the parameters in the deviance: the better the model fits the data, the larger are the values assumed by the likelihood function, and thus the smaller is $D(\langle \boldsymbol{\theta}\rangle_\mathrm{post})$. 
On the other hand, by adding $2p_D$ we penalise increasing model complexity  in order to avoid overfitting. Note that $\langle D\rangle_\mathrm{post}=D(\langle \boldsymbol{\theta}\rangle_\mathrm{post})+p_D$ already incorporates a penalty for complexity and should then be considered a measure of model adequacy rather than a pure measure of fit. The DIC can be interpreted as the Bayesian generalisation of the Akaike Information Criterion (AIC) used in the maximum-likelihood framework \cite{Akaike1998}. For non-hierarchical models and large data samples, the DIC asymptotically reduces to the AIC.
For a number of reasons\footnote{The concept of effective number of degrees of freedom was originally introduced to deal with hierarchical Bayesian models. In this case, the parameters that regulate observations at the individual level depend on hyperparameters (which are assigned hyperpriors) that describe the group level. In complex, multi-level hierarchies, parameters are not independent and it is not obvious how to calculate their total number. The advantage of introducing the $p_D$ estimator is that it uses MCMC results directly and straightforwardly.
In general, the $p_D$ statistic measures the constraining power of the data compared to the prior. 
Whenever the deviance is well approximated by a quadratic function around $\langle \boldsymbol{\theta}\rangle_\mathrm{post}$ (\ie the likelihood of the model parameters is approximately Gaussian as expected for large datasets from the Bayesian central limit theorem), each model parameter contributes one to $p_D$ if the posterior information about the parameter is dominated by the likelihood, it contributes zero if the information is dominated by the prior, and it contributes a number in between zero and one if both the prior and the likelihood are important to constrain its value \citep{GelmanEtal2004}. However, in peculiar cases in which the Gaussian approximation for the likelihood fails, $p_D$ could even assume negative values. Moreover, $p_D$ (and thus the DIC metric)  is not invariant under reparametrization of the model (while $\langle D \rangle_\mathrm{post}$ is).},
the definition of $p_D$ given above has been subject to much criticism.
An alternative estimator for the effective number of parameters in non-hierarchical models (which is invariant under reparametrization and never negative but only gives meaningful estimates when priors are non-informative) is $p_V=\mathrm{Var}(D)/2$ where $\mathrm{Var}(D)$ denotes the posterior variance of the deviance \cite{GelmanEtal2004}.  In this work, we use this second estimator to build the DIC metric from the MCMC runs.

The DIC statistic for a model is very easy to calculate when the likelihood is available in closed form and the posterior distribution is obtained by MCMC simulation. The actual value of the DIC for a model has no particular meaning, only differences $\Delta$DIC between models matter.
According to a commonly used rule of thumb, values of $\Delta \mathrm{DIC}<2$ are insignificant while differences of 5 and 10 provide substantial and decisive evidence against the less supported model. However, the reliability of this scale depends on the application. Tests have shown that the DIC tends to select overfitting models if $p_D$ is not small with respect to the sample size (\eg \cite{Plummer2008}).
Consistently, a number of theoretical studies suggest to increase the penalty for model complexity and use $\mathrm{DIC}^*=\langle D \rangle_\mathrm{post}+2p_D$
\cite{VanDerLinde2005, Plummer2008, VanDerLinde2012, Ando2011}.
In fact, under mild regularity assumptions, the original DIC selects the model that asymptotically (for large samples) gives the smallest expected Kullback-Leibler divergence between the DGP and the plug-in predictive distribution evaluated at $\langle \boldsymbol{\theta}\rangle_\mathrm{post}$. Instead, one would like to minimise the divergence with respect to 
the proper predictive distribution given in equation~(\ref{eq:predictivepdf}). 
We stress once again that the goal here is not to select the true model but rather to make a pragmatic choice that agrees with observations and provides good predictions for future datasets.

%%%%%%%%%%%%%%%%%%%%%%%%%%%%%%%%%%%%%%%%%%%%%%%%%%%%%%%%%%%%%%%%
\subsection{Goodness of fit and posterior predictive $p$-values}
\label{sec:ppp}

We would like to assess which models provide a good fit to the data and which do not. In classical statistics, the maximum-likelihood method is often used to determine the best-fit model parameters $\hat{\boldsymbol{\theta}}$ for a given dataset.
A test statistic (\eg the $\chi^2$) is then selected to determine the ``significance'' of the fit \cite{Fisher1925}. Under the null hypothesis that the data are actually sampled from the model with $\hat{\boldsymbol{\theta}}$, one computes the conditional frequentist probability of obtaining as many or more extreme data of the test statistic. This $p$-value corresponds to the long-run frequency taken over the sampling distribution of the data under the null hypothesis (\ie it is the probability that other unobserved data-sets would be more extreme than the one that was observed in terms of the test statistic).
When the $p$-value for an experiment is small, then one has to assume that either an unusual event has occurred or that the null hypothesis is not true. Thus, the smaller the $p$ value, the less it is plausible that the null hypothesis is true. 

Concern about the interpretation of $p$-values is widespread in statistics. Well-known problems of this approach are (i) that it is not possible to consider nuisance parameters and (ii) 
that it depends on how the data acquisition process is terminated and thus violates the likelihood principle.

Several authors have developed a Bayesian motivated adaptation of the classical goodness-of-fit test based on the $p$-value  \citep{Rubin1984, Meng1994, GelmanEtal1996, GelmanEtal2004}. 
The method relies on the posterior predictive probability distribution function given in equation~(\ref{eq:predictivepdf}).
This function gives the probability (conditional on the observed data) of replicated data that could have been observed or, to think in predictive terms, that would be observed in the future if the experiment is repeated. Note that what is kept fixed here is the observed data while the classical method
relies on probabilities that are conditional
%conditions 
on the parameters of the best-fit model $\hat{\boldsymbol{\theta}}$ (\ie the set $\hat{\boldsymbol{\theta}}$ is kept fixed).
The argument then proceeds as follows. A discrepancy variable $\Delta(\mathrm{data},\boldsymbol{\theta})$  is introduced to quantify the deviation of the model (with parameters $\boldsymbol{\theta}$) from the data. The posterior predictive $p$-value (ppp) of $\Delta$ is defined as
\be
    \mathrm{ppp}=\mathrm{Prob}\left[\Delta(\mathrm{replicated\ data},\boldsymbol{\theta})\geq \Delta(\mathrm{data},\boldsymbol{\theta}) | \mathrm{data}\right]\;,
\ee
where the probability is taken over the joint distribution 
\be
    P(\boldsymbol{\theta},\mathrm{replicated\ data}| \mathrm{data})=p(\mathrm{replicated\ data}|\boldsymbol{\theta})\, P(\boldsymbol{\theta}| \mathrm{data})\;.
\ee
In practice, we compute the ppp of $\Delta$ using MCMC simulations by drawing one replica from the statistical model for each step of the chain. The estimated ppp corresponds then to the fraction of steps for which the discrepancy variable equals or exceeds its realised value.
A ppp which is close to zero or one indicates that the realised data have a low probability of occurring under the postulated model, \ie that the model does not fit the data well. 
It is important to stress, however, that ppp's do not have in general a uniform distribution under the true model (meaning that the probability to find ppp $>0.95$ is not necessarily 5 percent) as they often tend to have distributions that are more concentrated around 0.5. Therefore, if one wants to associate a precise statistical significance to them, they need to be calibrated. 

In this work, we use the ppp as a measure of goodness of fit by adopting the log likelihood as the discrepancy variable. 
We perform the calibration of the ppp's by generating artificial data based on our reference model with added noise and by fitting them. We find that the distribution of our ppp's is remarkably close to uniform under the true model. Therefore, in the analysis of the N-body data, we interpret
extreme values of ppp near zero or one as revealing a systematic misfit between the bispectrum measurements and the model predictions that cannot be ignored.
However, in order to facilitate understanding to readers who are more familiar with frequentist goodness-of-fit tests, we also provide the value of the posterior averaged $\chi^2$ statistic and the corresponding upper one-sided 95 per cent confidence limit as a function of $\kmax$ (see Section \ref{sec:benchmark} for further details).
Note that the number of degrees of freedom that should be associated to this statistic is the total number of data points (as taking the posterior average gives a larger value than minimizing the $\chi^2$ as in equation (\ref{eq:dicpd}) for the deviance).

Finally, in order to quantify the degree to which a model systematically deviates from the actual measurements at the level of single data points, we make use of a technique known as graphical posterior predictive checking \citep{Rubin1984, GelmanEtal1996}.
In line with the ppp, this concept adopts a frequentist-like approach in a fully Bayesian framework.
The underlying idea is that, if a model provides a good fit, it could be used to generate replicated data that look like those that have been observed. In practice, we compute the difference between each realised data point (\ie the measurements from the N-body simulations) and the mean of the replicated data sampled from the posterior probability distribution function (\ie by considering one replica for each model sampled by the Markov chain). We then convert this difference into a standardized residual, $R$, by expressing it in units of the rms uncertainty of the data. Systematic deviations with $|R|\gg 1$ indicate potential shortcomings of the model.

%%%%%%%%%%%%%%%%%%%%%%%%%%%%%%%%%%%%%%%%%%%%%%%%%%%%%%%%%%%%%%%%
%%%%%%%%%%%%%%%%%%%%%%%%%%%%%%%%%%%%%%%%%%%%%%%%%%%%%%%%%%%%%%%%
%%%%%%%%%%%%%%%%%%%%%%%%%%%%%%%%%%%%%%%%%%%%%%%%%%%%%%%%%%%%%%%%
\section{Results}
\label{sec:results}

In this section, we present the results obtained using all the tools described above to fit the halo bispectrum with different versions of the tree-level model.

%%%%%%%%%%%%%%%%%%%%%%%%%%%%%%%%%%%%%%%%%%%%%%%%%%%%%%%%%%%%%%%%
%%%%%%%%%%%%%%%%%%%%%%%%%%%%%%%%%%%%%%%%%%%%%%%%%%%%%%%%%%%%%%%%
\subsection{Benchmark analysis}
\label{sec:benchmark}
As a starting point, we fit model $\mathcal{M}_3$ to our set of bispectrum measurements using  the likelihood function given in \eq{eq:SellentinLikelihood}. Bin averages for the model predictions are evaluated exactly using \eq{eq:binnedbisp}. 
We refer to this combination of choices as our reference study.

\begin{figure}
    \centering
    \includegraphics[width=\textwidth]{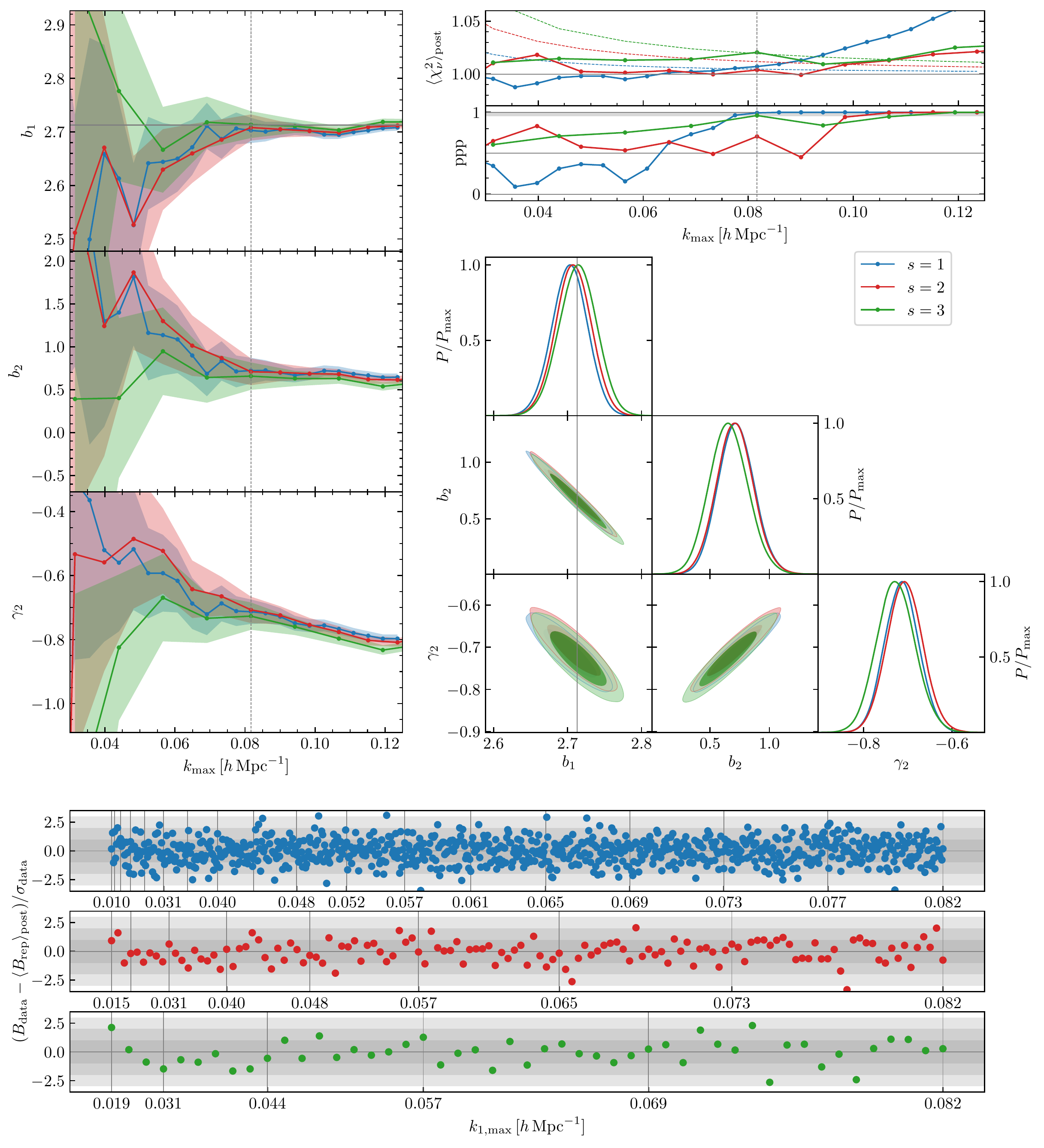}
    \caption{Fit of model $\mathcal{M}_3$ to the halo bispectrum measured using three different bin widths $\Delta k = k_f, 2k_f, 3k_f$. 
    The top-left panels show the posterior mean (solid lines) and rms scatter (shaded areas) of the model parameters as a function of $\kmax$. The vertical dashed line highlights the reference scale of $\kmax = 0.082 \kMpc$ for which we display contour plots for the joint posterior density of parameter pairs in the middle-right panel. Here, darker and lighter shaded areas represent the 68 and 95 percent joint credibility regions, respectively. The narrow gray bands indicate the constraints on the linear-bias parameters derived from the halo-matter cross power spectrum.
    The standardized residuals of the posterior predictive checks for the same $\kmax$ are shown in the large panel on the bottom.
    Two goodness-of-fit diagnostics are displayed in the top-right panel as function of $\kmax$: the reduced $\chi^2$ statistic averaged over the posterior (top inset) and the $ppp$ (bottom inset). As a reference, the dashed curves in the top inset indicate the upper one-sided 95 percent confidence limit in a frequentist $\chi^2$ test (note that the number of datapoints included in the fit varies with  $\kmax$).
    }
    \label{fig:benchmark3parsA}
\end{figure}

Figure~\ref{fig:benchmark3parsA}, like many others that follow, displays our results in five complementary panels.
In the top-left corner, we show the mean and the rms values of the posterior distribution\footnote{We do not correct the width of the posteriors for the loss of information due to the uncertainty in the covariance matrix \cite{PercivalEtal2014}. In fact, this is impossible to do in an exact way \cite{SellentinHeavens2017}. Moreover, for all scales at which we obtain a good fit, the expected size of the correction (a few percent) is several times smaller than the statistical uncertainty with which we measure the rms value of the model parameters in the MCMC simulations.} for each model parameter as a function of $\kmax$.
The goodness of fit is displayed in the top-right panel. Here, we show the ppp\footnote{Replicated data are generated by adding Gaussian noise (with covariance matrix $\Ce$) to the theoretical models at each step of the MCMC simulations, meaning that the quantity $\delta {\vec B}_\mathrm{repl} \cdot \Ce^{-1} \cdot \delta {\vec B}_\mathrm{repl}$ follows a $\chi^2$ distribution with $N_\mathrm{t}$ degrees of freedom for each realisation.} and the posterior-averaged $\chi^2$ statistic divided by the number of data points, both as a function of $\kmax$. 
The central panel presents contour plots for the joint posterior density of all parameter pairs at $\kmax=0.082\kMpc$. This particular value has been selected for two reasons: i) it is an exact multiple of all the bin sizes we consider, thus ensuring that all measurements contain the same information; ii)  it approximately coincides with the largest $\kmax$ for which the fits we present are good.
Finally, in the bottom panel, we provide a direct comparison of the data and the fit (at $\kmax=0.082\kMpc$) in terms of the standardized residuals of the posterior predictive checks (PPCs) described at the end of section~\ref{sec:ppp}. In practice, we compute the difference between the measured bispectrum and the posterior mean of the replicas. The result is then averaged over all the realizations and normalized to the standard deviation of the measurements.
Equilateral (binned) configurations are highlighted with vertical lines, so that from one vertical line to the other we span all configuration shapes with $k_1\ge k_2\ge k_3$ at fixed $k_1$.

In order to test the reliability of the fits presented above,
we also derive the linear-bias parameter by directly comparing the halo-matter cross power spectrum and the matter power spectrum of the N-body simulations at very large scales. We find a value of $b_1^\times = 2.7133 \pm 0.0007$, where the superscript here indicates that this measurement comes from cross-correlating halos and matter. In figure~\ref{fig:benchmark3parsA}, this result is represented as a narrow horizontal band in the top-left panel and as a vertical band in the center-right panel, both painted gray.

Coming to the specific results of our benchmark study, we note that the range of validity of model $\mathcal{M}_3$ does not extend beyond $\kmax=0.1\kMpc$ and shows some dependence on the bin size. 
The ppp approaches one at $\kmax$ of, approximately, $0.082, 0.107$ and $0.119 \kMpc$ for $\Delta k = k_f, 2k_f$ and $3k_f$, respectively. Similar conclusions can be drawn based on $\langle \chi^2 \rangle_{\mathrm{post}}$.
The values of $\kmax$ for which a good fit is achieved also correspond to consistent posterior probability distributions for the model parameters. 
All fits provide compatible results (within the statistical uncertainty) up to $\kmax\sim 0.08\kMpc$. When we include smaller scales in the analysis, instead, the location of the posterior for $\gamma_2$ starts running with $\kmax$ and we simultaneously obtain unsatisfactory values of the goodness-of-fit statistics.
The posterior probabilities for $\kmax=0.082 \kMpc$ show marked degeneracies between the model parameters that cannot be individually constrained apart from $b_1$. The standardized residuals of the PPCs do not single out any particular triangle configuration for which the model systematically deviates from the data for any binning scheme. Note that the goodness of fit at large $\kmax$ could deteriorate either because the tree-level models become inaccurate on small scales but also because our estimate of the covariance matrix becomes more and more imprecise.

The relatively small difference among the results for different bin sizes are to some extent expected. In general, the ability of the bispectrum to break degeneracies between parameters depend on the different dependence on shape of the different contributions. This is clearly affected by the binning. In particular, it is reduced by a larger binning. At the same time, a larger binning also implies smaller error bars. The combination of such different factors is likely at the origin of the observed differences.

%%%%%%%%%%%%%%%%%%%%%%%%%%%%%%%%%%%%%%%%%%%%%%%%%%%%%%%%%%%%%%%%
%%%%%%%%%%%%%%%%%%%%%%%%%%%%%%%%%%%%%%%%%%%%%%%%%%%%%%%%%%%%%%%%
\subsection{Model selection: shot noise}
\label{ssec:model-shotnoise}
\begin{figure}
    \centering
    \includegraphics[width=\textwidth]{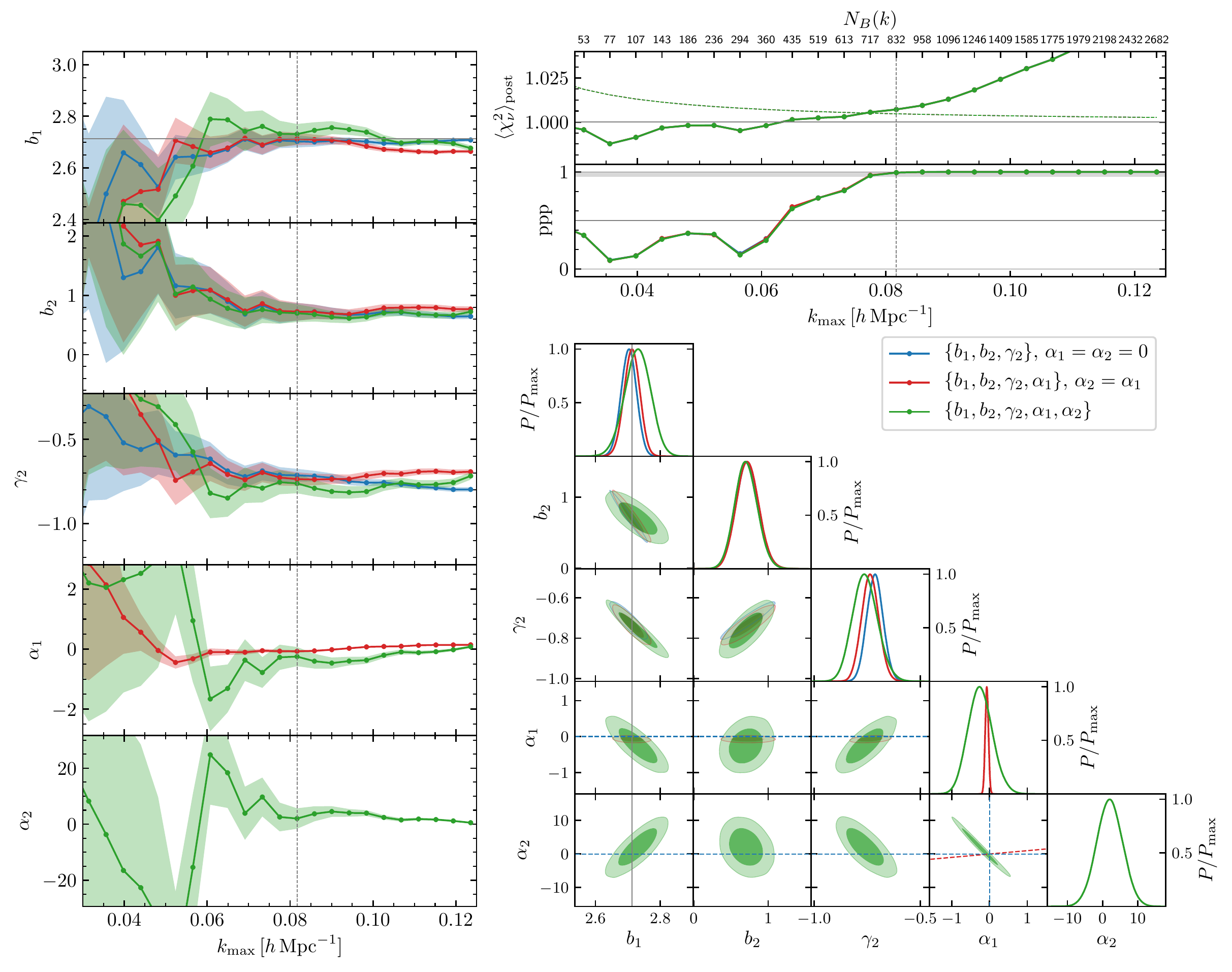}
    \caption{As in figure~\ref{fig:benchmark3parsA} but comparing models $\mathcal{M}_3$, $\mathcal{M}_4$ and $\mathcal{M}_5$ for     the binned data with $s=1$ and the broad priors for $\alpha_1$ and $\alpha_2$. The labels in the inset summarize the difference between the models: the parameters between curly brackets are let free to vary within the prior range while the others are kept fixed. The blue dashed lines show the relations $\alpha_1 = \alpha_2 = 0$ in $\mathcal{M}_3$, while the red dashed line shows the relation $\alpha_2 = \alpha_1$ in $\mathcal{M}_4$.}
    \label{fig:modelcomparisondk1}
\end{figure}

We now fit models ${\mathcal M}_4$ and ${\mathcal M}_5$ to the bispectrum data. They
extend model ${\mathcal M}_3$ to include corrections to Poissonian shot noise that are routinely included in the analysis of survey data, \eg \cite{GilMarinEtal2017, CastorinaEtal2019}. 
Figure \ref{fig:modelcomparisondk1} compares the results for the three models for the measurements with $s=1$ when using broad priors for the shot-noise parameters. Analogous results are obtained adopting one of the other binning schemes.
The most important thing to notice is that adding extra shot-noise parameters does not lead to any significant improvement in the goodness of fit and thus does not extend the range of validity of the models in terms of $\kmax$. 
Further insight is obtained by looking at the PPCs for $\kmax=0.082 \kMpc$ (not shown): the residuals are virtually identical for the three models.
In general, the fit results for ${\mathcal M}_4$ are very similar to ${\mathcal M}_3$ as, for sufficiently large $\kmax$, the additional parameter is always well constrained to be close to zero. On the other hand, large degeneracies between $\alpha_1$ and $\alpha_2$ (that, in this case, can assume opposite signs) as well as between the shot-noise and the bias parameters are present for ${\mathcal M}_5$. This leads to substantially larger marginalised posteriors for the bias parameters. It is also worth stressing that the loose constraints set by the data on $\alpha_1$ and $\alpha_2$ span a much larger range than expected from theoretical shot-noise models \cite{BaldaufEtal2013, GinzburgDesjacquesChan2017}. We thus conclude that the large-scale bispectrum is insufficient to inform these models.

\begin{figure}
    \centering
    \includegraphics[width=\textwidth]{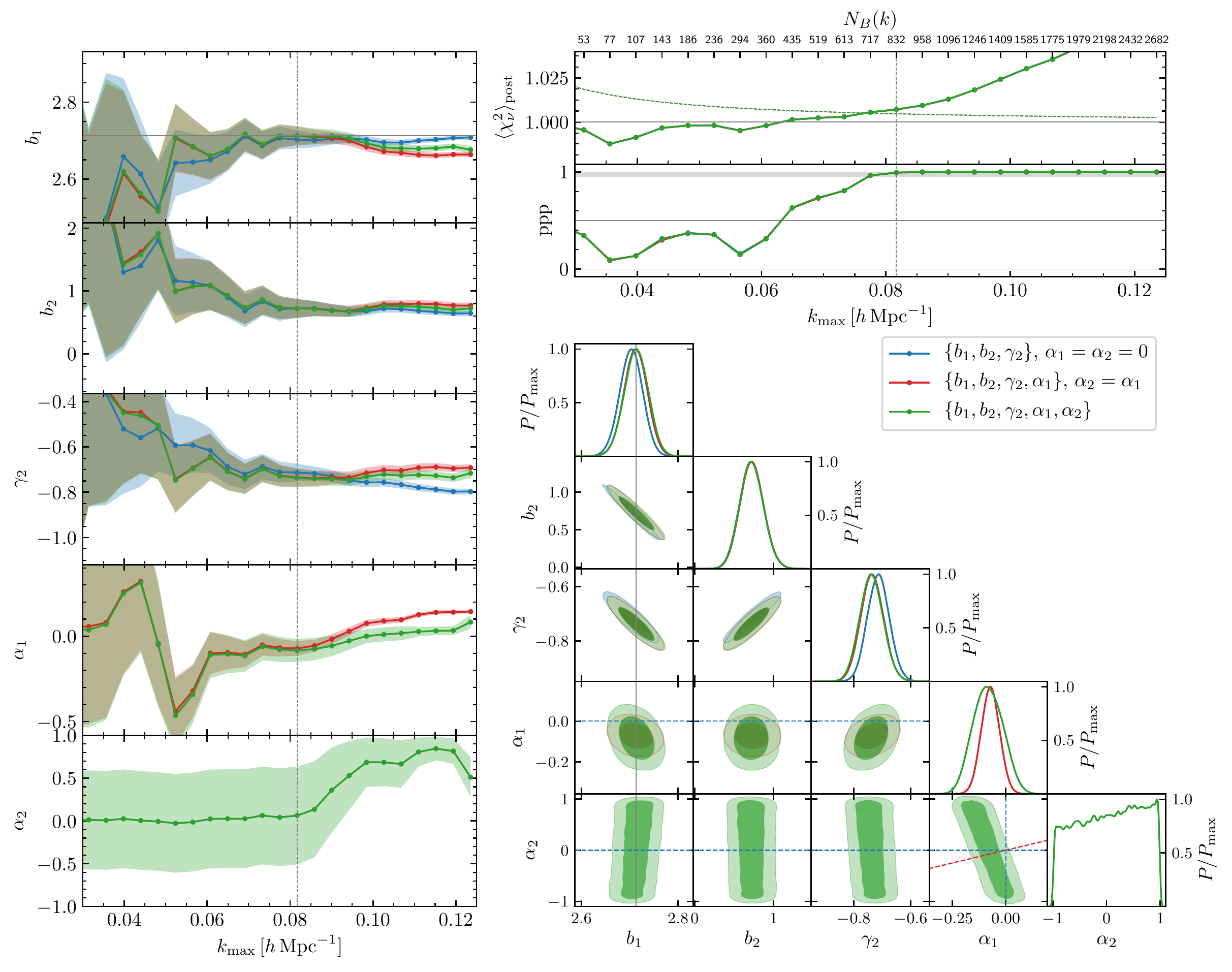}
    \caption{Same as figure~\ref{fig:modelcomparisondk1} but assuming narrow priors on the shot-noise parameters, \ie $\alpha_{1,2} \in [-1,1]$. While $\alpha_2$ is completely unconstrained where the model still gives a good fit to the data, the posteriors for the other parameters do not change sensibly when adding parameters.}
    \label{fig:modelcomparisondk1v2}
\end{figure}

Figure \ref{fig:modelcomparisondk1v2} shows a comparison similar to the one shown in figure \ref{fig:modelcomparisondk1}, but for the narrower priors on the shot-noise parameters. As in the previous case, both for $\mathcal M_4$ and $\mathcal M_5$, $\alpha_1$ is well constrained and consistent with 0, while in $\mathcal M_5$, $\alpha_2$ is completely unconstrained. On the other hand, the results for the bias parameters are essentially unaffected by the introduction of shot-noise parameters with such priors.

\begin{figure}
    \centering
    \includegraphics[width=0.9\textwidth]{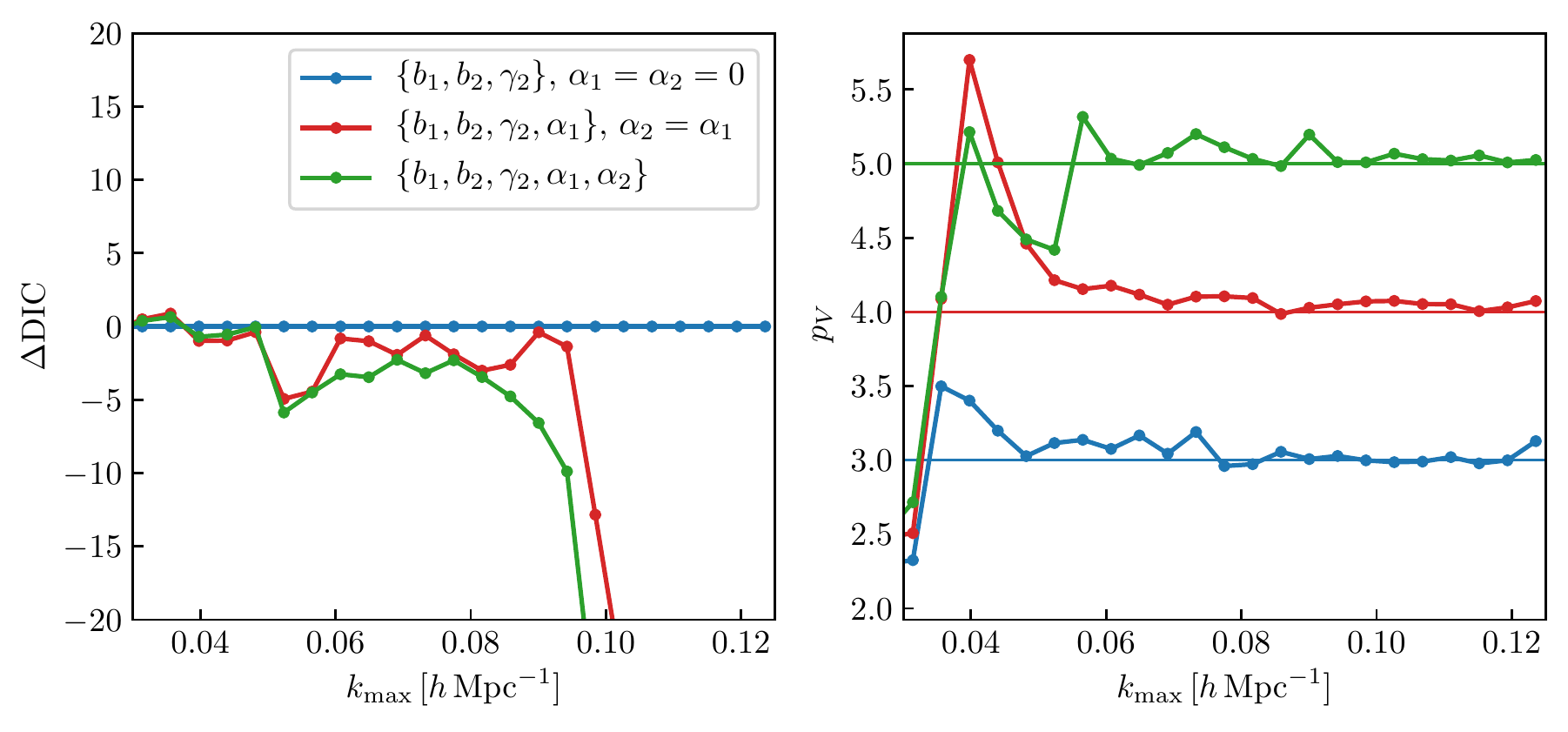}
    \includegraphics[width=0.9\textwidth]{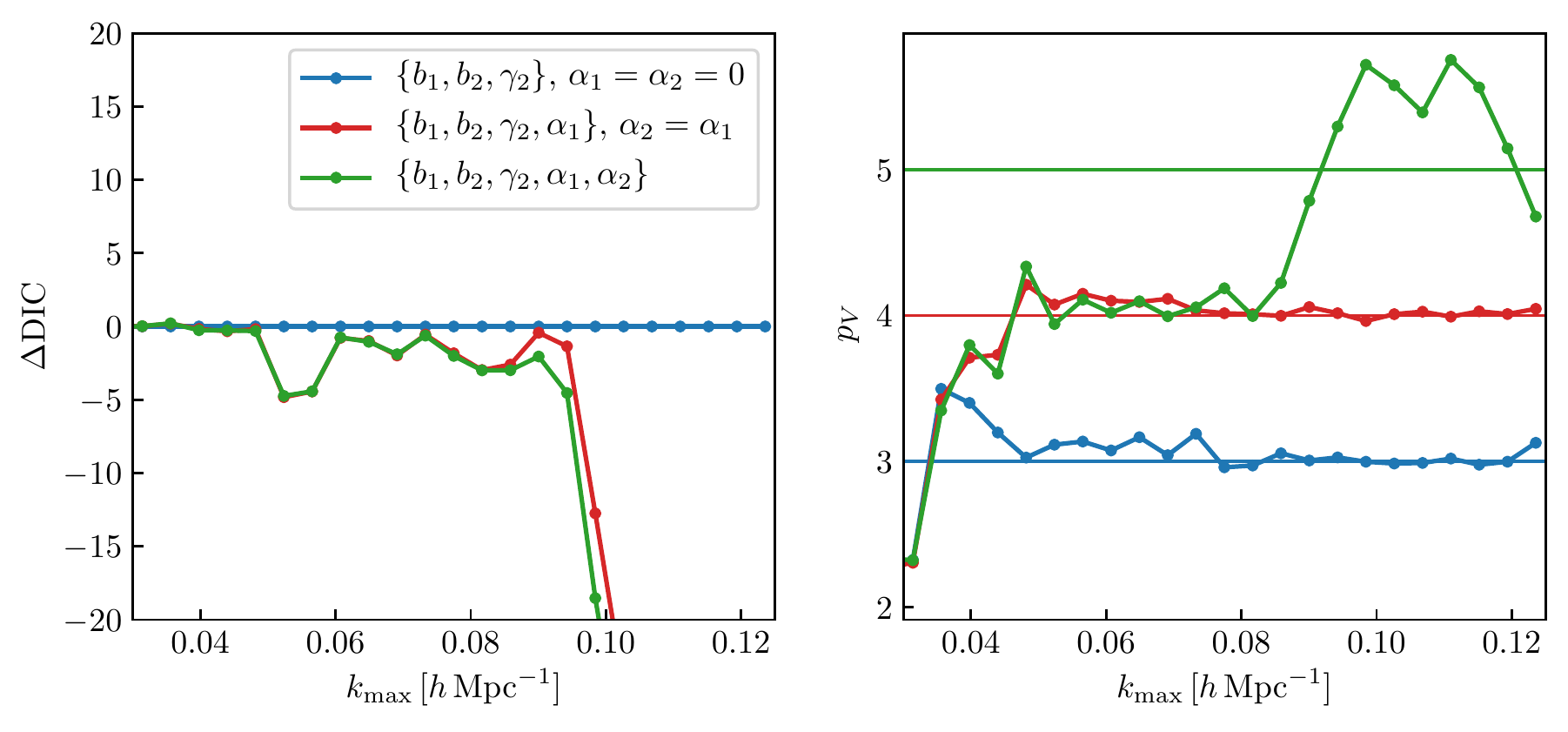}
    \includegraphics[width=0.45\textwidth]{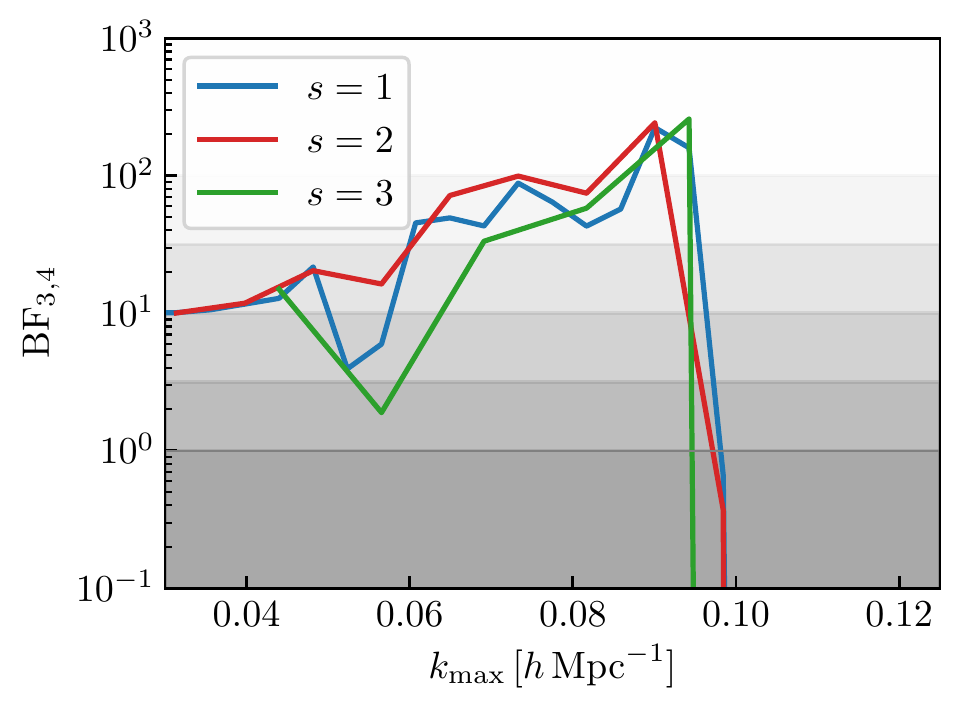}
    \includegraphics[width=0.45\textwidth]{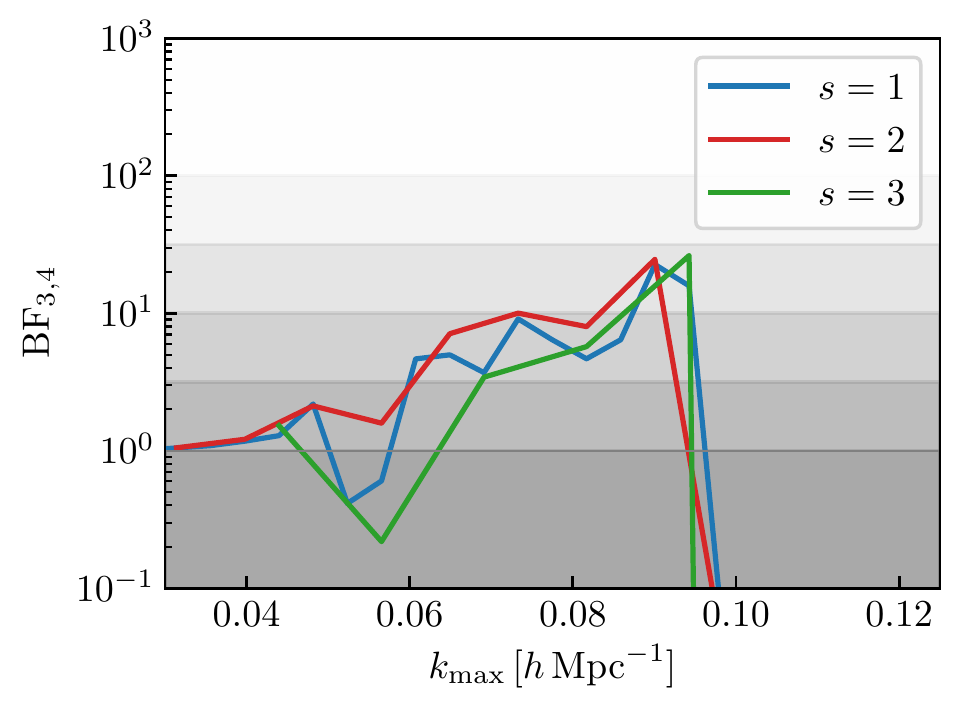}
    \caption{Model comparison between the fits presented in figure~\ref{fig:modelcomparisondk1} and figure~\ref{fig:modelcomparisondk1v2}. The DIC difference with respect to model $\mathcal{M}_3$ is presented in the top-left panel as a function of $\kmax$ and the corresponding effective number of parameters $p_V$ is displayed in the top-right panel for the case of broad priors, while corresponding quantities for narrow shot-noise priors are shown in the middle-left and middle-right panels.
    The bottom panels show the BF from the Savage-Dickey density ratio for the comparison between $\mathcal{M}_3$ and $\mathcal{M}_4$ for broad priors (bottom-left) and for narrow priors (bottom-right).
    In this case, results are shown for the three different bin widths, $s = 1, 2$ and $3$
    (solid blue, red and green respectively).
    The shaded areas represent the Jeffreys' scale for the BF and correspond to regions where the evidence for model $i$ over model $j$ is, from bottom to top, ``negative'', ``barely worth mentioning'', ``substantial'', ``strong'', ``very strong'' and ``decisive''. Model $\mathcal M_3$ appears to be preferred over $\mathcal M_4$ and $\mathcal M_5$ according to both the model selection tests ($\rm BF_{4,5}$ not shown). Note that, while the DIC is slightly larger in the case of $\mathcal M_3$, the difference is not so large to favour the other models (that also have a larger number of parameters).}
    \label{fig:DIC_SD}
\end{figure}

In figure~\ref{fig:DIC_SD}, we apply model-selection techniques to the nested models $\mathcal{M}_3$, $\mathcal{M}_4$, and $\mathcal{M}_5$ considering both priors sets.
We first look at the constraining power of our bispectrum data with $s=1$ (similar results are found for the other bin sizes).
The top-right and middle-right panels show the effective number of parameters of the fits as a function of $\kmax$, respectively for the broad and narrow shot-noise priors. While all parameters of $\mathcal{M}_3$ and $\mathcal{M}_4$ stop being prior dominated for $\kmax \simeq 0.04 \kMpc$, the bispectrum data can fully constrain $\mathcal{M}_5$ only from $\kmax \simeq 0.06 \kMpc$ for the broad priors, and from $\kmax \simeq 0.08 \kMpc$ for the narrow priors.
In the top-left and middle-left panels, instead, we show the DIC difference with respect to $\mathcal{M}_3$, again for broad and narrow priors respectively.
In both cases, for small values of $\kmax$, $\mathcal{M}_3$ is slightly preferred by the DIC, although with low significance. A fair  conclusion is that, for $\kmax<0.09 \kMpc$, the three models provide very similar DIC and cannot be ranked. On smaller scales, the DIC strongly prefers $\mathcal{M}_5$ but this is irrelevant as none of the models provides an acceptable goodness of fit. It is, in fact, quite possible that some non-linear effects that are not included in the tree-level model are partially accounted for by the shot-noise corrections. 
Stronger conclusions can be drawn based on the Bayes factors evaluated using the Savage-Dickey density ratio (bottom panels). In this case, for all binning schemes, $\mathcal{M}_3$ is decisively preferred over $\mathcal{M}_4$ at $\kmax\simeq 0.09\kMpc$ where the ratio exceeds 100 with broad shot-noise priors, and 10 for narrow shot-noise priors (although it suddenly drops for larger values of $\kmax$ where both models fail to properly fit the data). 
Similarly, $\mathcal{M}_4$ is preferred over $\mathcal{M}_5$ although to a lesser degree as the BF (not shown) only reaches a maximum value of about 20 at $\kmax\simeq 0.08\kMpc$. The bottom line of this section is that our data provide no evidence for non-Poissonian shot-noise corrections on the scales in which the tree-level model fits well, \ie $\kmax<0.08\kMpc$.

%%%%%%%%%%%%%%%%%%%%%%%%%%%%%%%%%%%%%%%%%%%%%%%%%%%%%%%%%%%%%%%%
%%%%%%%%%%%%%%%%%%%%%%%%%%%%%%%%%%%%%%%%%%%%%%%%%%%%%%%%%%%%%%%%
\subsection{Model selection: reducing the number of bias parameters}
\label{ssec:model-bias}
In section~\ref{sec:theomod}, we have described several possibilities for reducing the freedom of our reference theoretical model $\mathcal{M}_3$.
We now investigate whether these restricted models provide an accurate description of the bispectrum measurements extracted from our simulations and contrast them with $\mathcal{M}_3$.

\begin{figure}[t]
    \centering
    \includegraphics[width=0.95\textwidth]{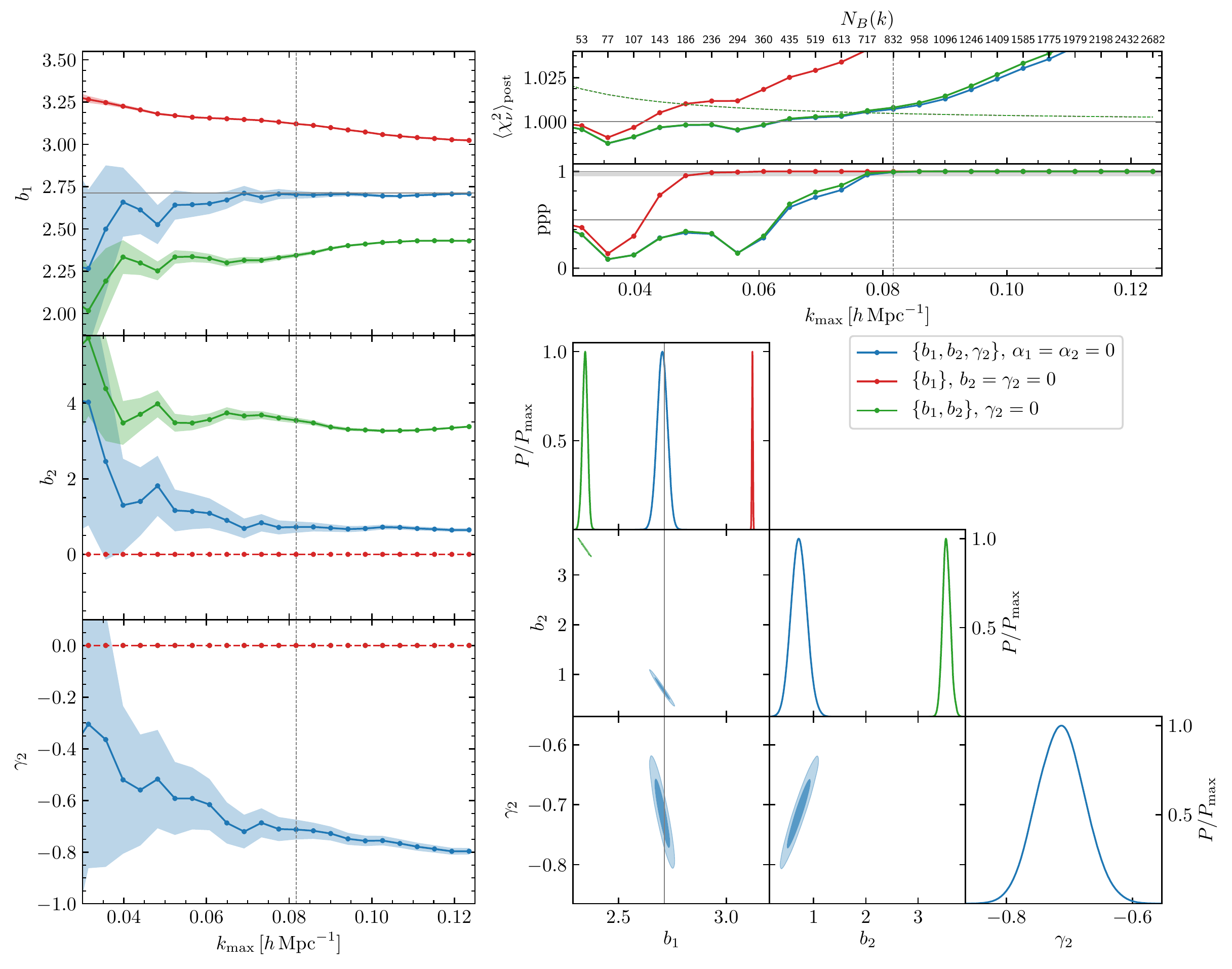}\\
    \includegraphics[width=0.88\textwidth]{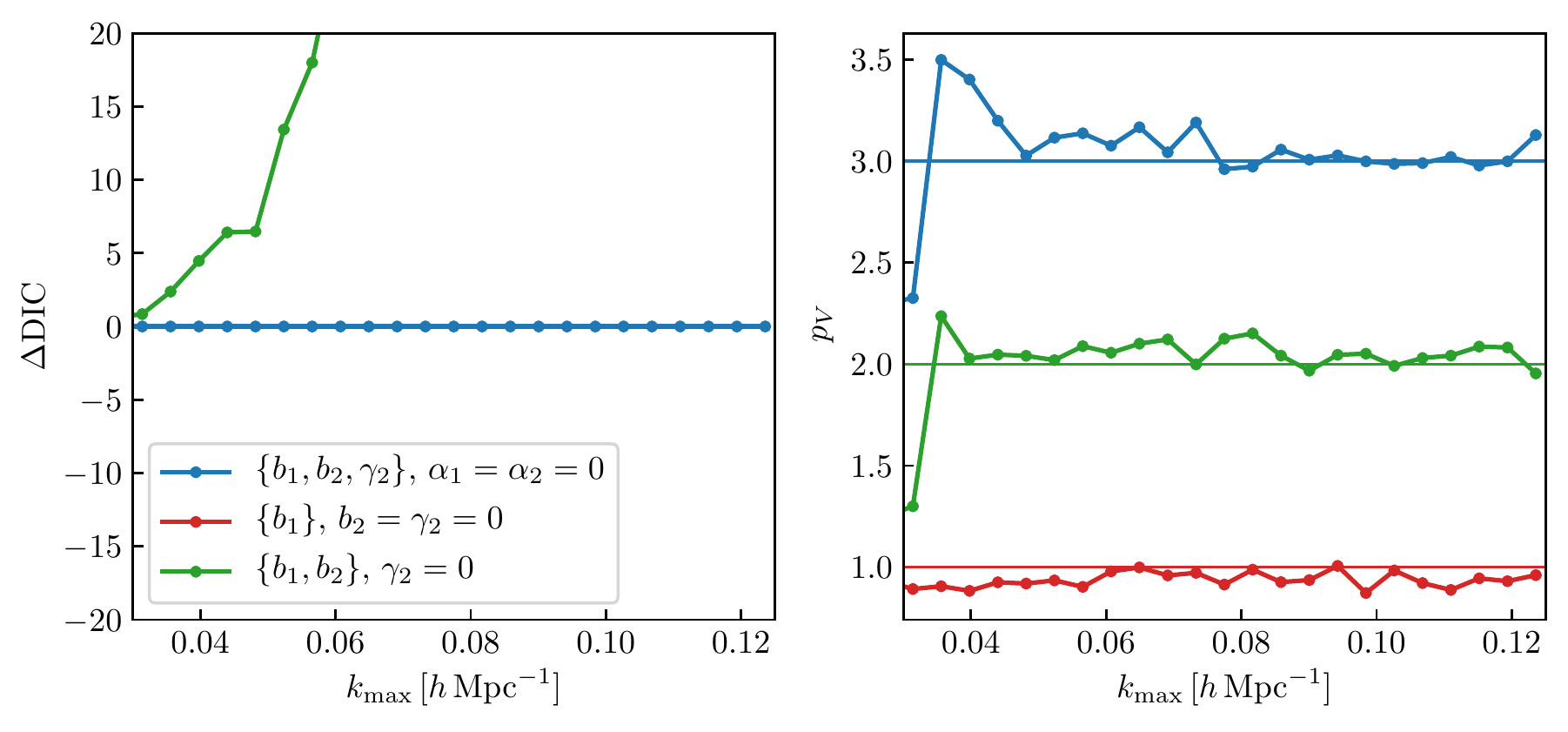}
    \caption{As in figures~\ref{fig:modelcomparisondk1} and \ref{fig:DIC_SD} but now comparing models $\mathcal{M}_1$, $\mathcal{M}_{2\mathrm{loc}}$ and $\mathcal{M}_3$. With respect to $\mathcal M_3$, both $\mathcal M_1$ and $\mathcal M_{2 \rm loc}$ give biased values for the parameters that have been left free to vary; the data clearly prefers $\mathcal M_3$ over the other two models.}
    \label{fig:modelcomparisondk1_local}
\end{figure}

As a first test, in figure~\ref{fig:modelcomparisondk1_local}, we compare models $\mathcal{M}_1$ and $\mathcal{M}_{2\mathrm{loc}}$ (which correspond to truncating a local Eulerian bias expansion at first and second order, respectively) with $\mathcal{M}_{3}$ (which also includes the tidal-bias term). In terms of goodness of fit, $\mathcal{M}_1$ fails around $\kmax\simeq 0.05\kMpc$, \ie at significantly larger scales than the other models that, on the other hand, provide almost identical values for the ppp and $\langle \chi^2_\nu \rangle$.
The local models retrieve different values for the bias parameters with respect to $\mathcal{M}_3$ (see also \cite{ChanScoccimarroSheth2012, BaldaufEtal2012}). 
Without combining our results with other clustering statistics, it is impossible to say whether $\mathcal M_{2\rm loc}$ and $\mathcal M_3$ provide a realistic description of the data since, in terms of goodness of fit, they are practically equivalent.
However, it is interesting to notice that the DIC shown in the bottom-left panel of figure~\ref{fig:modelcomparisondk1_local} not only strongly disfavours $\mathcal{M}_1$ already at $\kmax\simeq 0.03\kMpc$, but also clearly indicates a preference for $\mathcal{M}_3$ at $\kmax\simeq 0.06\kMpc$. Equivalent conclusions can be drawn from the Savage-Dickey ratios (not shown in the figure for the sake of brevity).

\begin{figure}[t]
    \centering
    \includegraphics[width=0.95\textwidth]{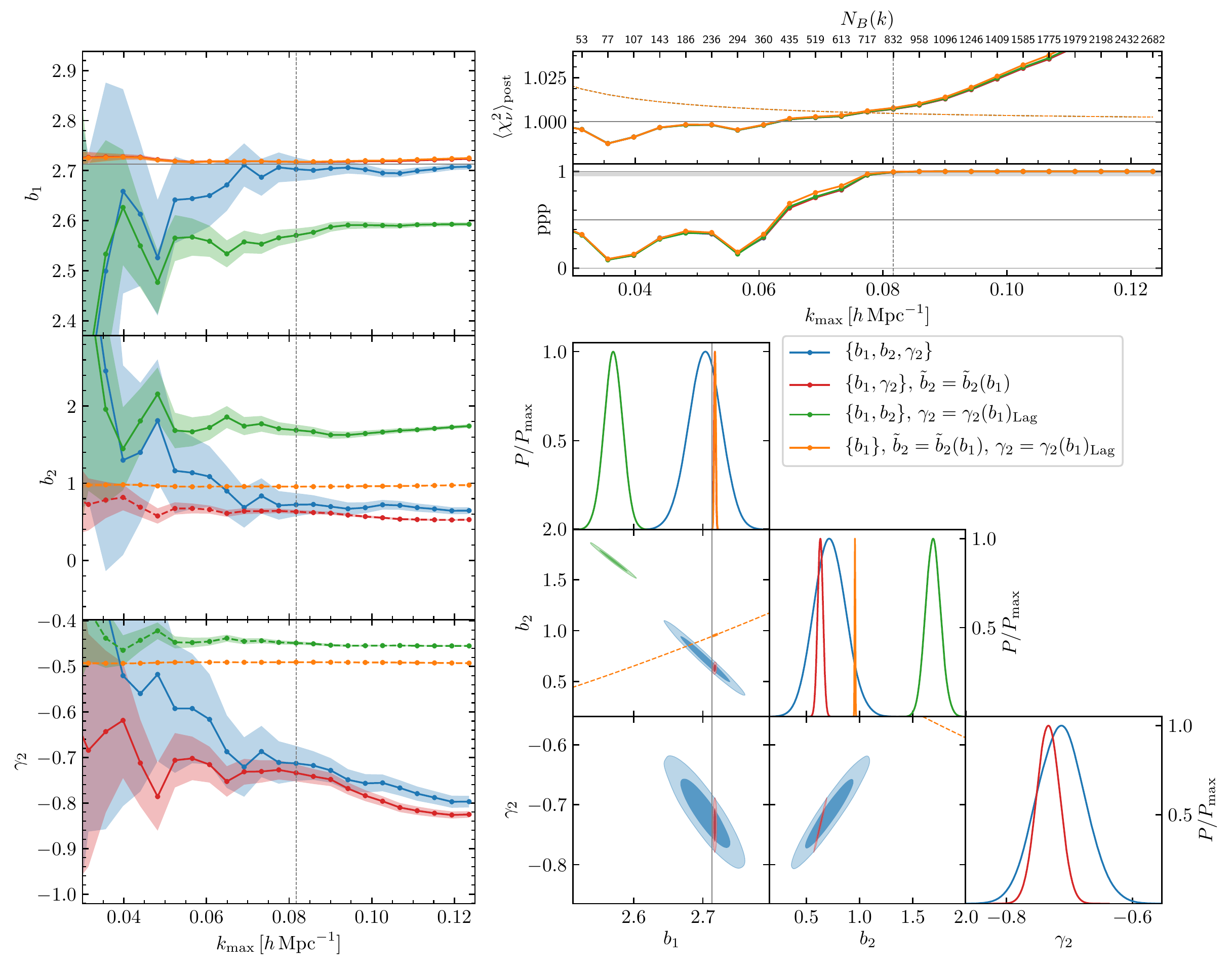}
    \includegraphics[width=0.88\textwidth]{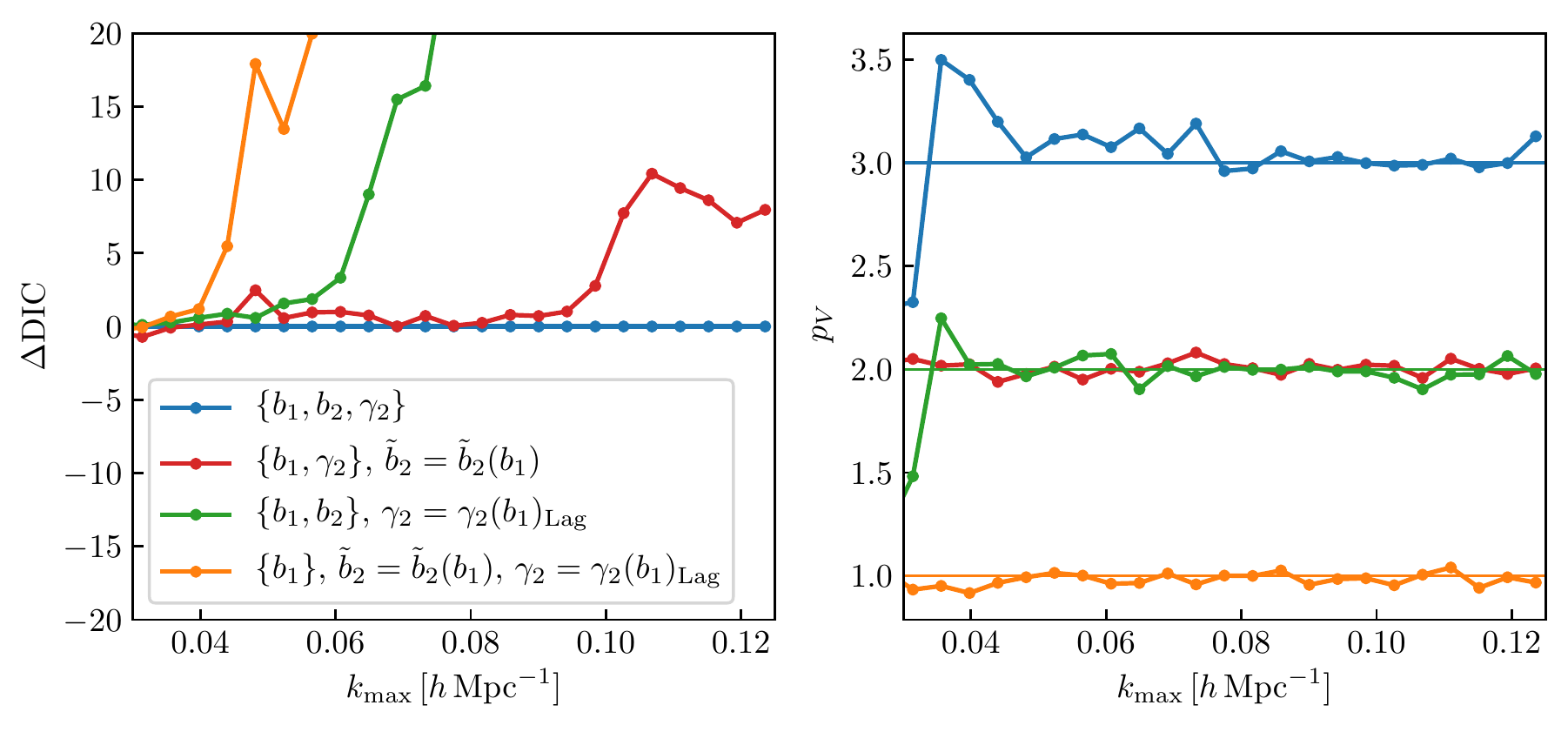}
    \caption{As in figures~\ref{fig:modelcomparisondk1} and \ref{fig:DIC_SD} but now comparing models $\mathcal{M}_{1 f}$, $\mathcal{M}_{2b_2}$, $\mathcal{M}_{2\gamma_2}$ and $\mathcal{M}_3$. 
    The orange dashed lines in the contour plot represent the combination of the fitting function $\tilde{b}_2(b_1)$ presented in \cite{LazeyrasEtal2016} and the relation $\gamma_2(b_1)$ derived from the assumption of local-Lagrangian biasing. Model $\mathcal M_{2 b_2}$ gives biased values for the fitted parameters as a function of $k_{\rm max}$ with respect to $\mathcal M_3$; while model $\mathcal M_{1 f}$ gives values of $b_1$ consistent with respect to the ones found with $\mathcal M_3$, the values of the derived parameters are biased with respect to $\mathcal M_3$; model $\mathcal M_{2 \gamma_2}$ appears to be consistent with $\mathcal M_3$, in posteriors for the parameters and also according to the DIC. Both $\mathcal M_{1 f}$ and $\mathcal M_{2 b_2}$ are disfavoured according to the DIC.}
    \label{fig:modelcomparisondk1_bias}
\end{figure}

A second series of tests is conducted in figure~\ref{fig:modelcomparisondk1_bias}. Here, we consider the models $\mathcal{M}_{1f}$, $\mathcal{M}_{2b_2}$ and $\mathcal{M}_{2\gamma_2}$ that have been obtained by mathematically relating the bias parameters of $\mathcal{M}_3$. 
It is important to remember that the relations we use have different origins.
While the function $\tilde{b}_2(b_1)$ is a fit to N-body simulations \cite{LazeyrasEtal2016}, $\gamma_2(b_1)$ embodies an assumption characterizing a class of bias models that have been already ruled out by recent studies \cite{LazeyrasSchmidt2018, AbidiBaldauf2018}.  
We find that all models achieve essentially the same goodness of fit as our benchmark model.
On the other hand, the posterior distributions of the parameters as a function of $\kmax$ show significant differences between $\mathcal{M}_3$ and all cases in which the local-Lagrangian relation for $\gamma_2$ is adopted. 
Only the results obtained with $\mathcal{M}_{2\gamma_2}$ are largely consistent with the reference model. One can also notice how assuming the $\tilde{b}_2(b_1)$ relation greatly reduces error bars with respect to what happens when assuming instead the $\gamma_2(b_1)$ relation. This is probably related to the fact that $b_1$ and $b_2$ are more strongly degenerate than $b_1$ and $\gamma_2$ in the fit for the benchmark model as evidenced in the contour plots at the reference scale.
Also note that, when we impose one single relation among the bias coefficients, the joint posteriors for the parameters, including the derived ones, are fully consistent with the results obtained for $\mathcal{M}_3$ at the reference scale. On the other hand, if both relations are imposed simultaneously, we find significant tension in the joint posteriors with respect to the benchmark model, thus suggesting that the local-Lagrangian approximation is not compatible with the $\tilde{b}_2(b_1)$ fit. In fact, the DIC clearly disfavours both $\mathcal{M}_{1f}$ and $\mathcal{M}_{2b_2}$ while $\mathcal{M}_{2\gamma_2}$ and $\mathcal{M}_{3}$ are essentially equivalent up to $0.1\kMpc$.
It is important to understand how the figure on the goodness of fit (in which all the models seem to fit the data equally well) can be reconciled with the conclusions we have drawn from the DIC. The key is to remember that we are fitting a very large number of data points, namely 247,936 at the reference scale of $\kmax=0.082 \kMpc$.
The DIC is driven by the fact that the posterior average of $-\log \mathcal{L}$ for $\mathcal{M}_3$ is smaller than that obtained with $\mathcal{M}_{1f}$ by nearly 18.
Since this difference is substantially larger than the number of extra parameters in $\mathcal{M}_3$, the DIC prefers this model.
On the other hand, the goodness-of-fit statistic $\langle \chi^2_\nu\rangle_{\mathrm{post}}$ essentially depends on the average log-likelihood per data point which is very similar for all models. 
Therefore, although all models considered in this section provide a statistically acceptable fit of the data, $\mathcal{M}_{2\gamma_2}$ and $\mathcal{M}_3$ are preferred as they better describe our ensemble of measurements on small scales and do not overfit.

In all fits we have analysed so far, the posterior distribution of $b_1$ was always in good agreement with the measurement of $b_1 ^\times$.
However, this is not the case for many of the models presented in figure~\ref{fig:modelcomparisondk1_bias}. For instance, $\mathcal M_{2b_2}$ gives a strongly biased estimate of $b_1$ (with respect to $b_1 ^\times$) for wavenumbers larger than $\sim 0.06 \kMpc$.
In addition, while the posteriors of $b_1$ from $\mathcal M_{2 \gamma_2}$ and $\mathcal M_{1 f}$ appear to be closer to the measured value $b_1^\times$, both of them are actually more than $3 \sigma$ away from $b_1 ^\times$. Therefore, of the four models analysed here, only our benchmark model $\mathcal M_3$ gives values of $b_1$ that are fully consistent with $b_1 ^\times$, and thus it is the only model giving fully unbiased values of the parameters. The model-selection techniques we implemented allowed us to single out the reference model without prior knowledge of the actual values of the parameters.
We envision that model-selection diagnostics will be particularly useful when considering more complex theoretical models that include bias loop corrections and depend on a much larger number of parameters.

%%%%%%%%%%%%%%%%%%%%%%%%%%%%%%%%%%%%%%%%%%%%%%%%%%%%%%%%%%%%%%%%
%%%%%%%%%%%%%%%%%%%%%%%%%%%%%%%%%%%%%%%%%%%%%%%%%%%%%%%%%%%%%%%%
\subsection{Binning of theoretical predictions}
\label{ssec:binning}

\begin{figure}[t]
    \centering
    \includegraphics[width=\textwidth]{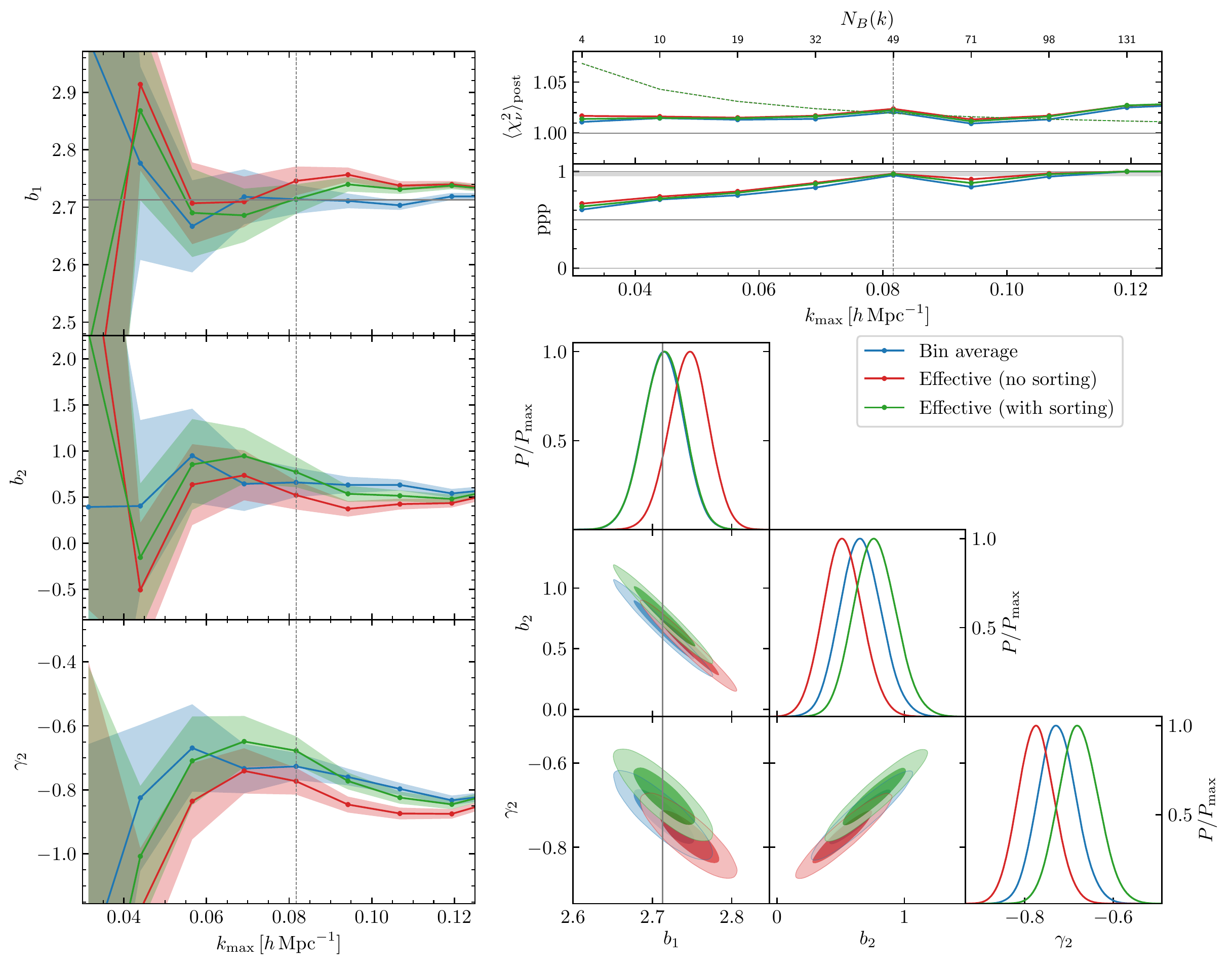}
    \caption{As in figure~\ref{fig:modelcomparisondk1} with $\Delta k=3 k_f$ but now comparing the different methods to evaluate the theoretical model for the binned bispectrum presented in section~\ref{ssec:binningtheory} and figure~\ref{fig:prediction_comp}. While the goodness-of-fit tests do not clearly prefer any of the different methods, the effective approaches give slightly biased values for the parameters, but still consistent with the ones found with a full bin average of the theoretical prediction.}
    \label{fig:effectivedk3}
\end{figure}

All results presented so far rely on averaging the models for the bispectrum over all fundamental triangles that correspond to a given triangle bin following \eq{eq:binnedbisp}. In this section, we investigate the impact of using simpler but less accurate theoretical predictions that require a single evaluation of the bispectrum at a triplet of effective wavenumbers defined either as in \eq{eq:effectiveks1} or (\ref{eq:effectiveks2}). 
Figure~\ref{fig:effectivedk3} shows the influence of the different methods on the fit of model $\mathcal{M}_3$ to the bispectrum data with $\Delta k = 3\,k_f$, for which we expect the largest variations.
For the $\kmax$ range in which the fit is good, 
the posteriors for $b_1$ obtained with the sorted effective wavenumbers lead to rather small differences with those obtained with exact binning, while $b_2$ and $\gamma_2$ show a deviation of $\sim 0.5 \sigma$; using the unsorted effective wavenumbers introduces a slightly larger bias in all parameters.

We are not showing here the results corresponding to theoretical predictions evaluated on the triangle bin centers as this approach applies only ``closed'' triangle bins. A comparison limited to such configurations would show, as we can expect, substantially biased estimates for the model parameters.

%%%%%%%%%%%%%%%%%%%%%%%%%%%%%%%%%%%%%%%%%%%%%%%%%%%%%%%%%%%%%%%%
%%%%%%%%%%%%%%%%%%%%%%%%%%%%%%%%%%%%%%%%%%%%%%%%%%%%%%%%%%%%%%%%
\subsection{Likelihood function}
\label{ssec:CovStatUnc}

\begin{figure}[t]
    \centering
    \includegraphics[width=\textwidth]{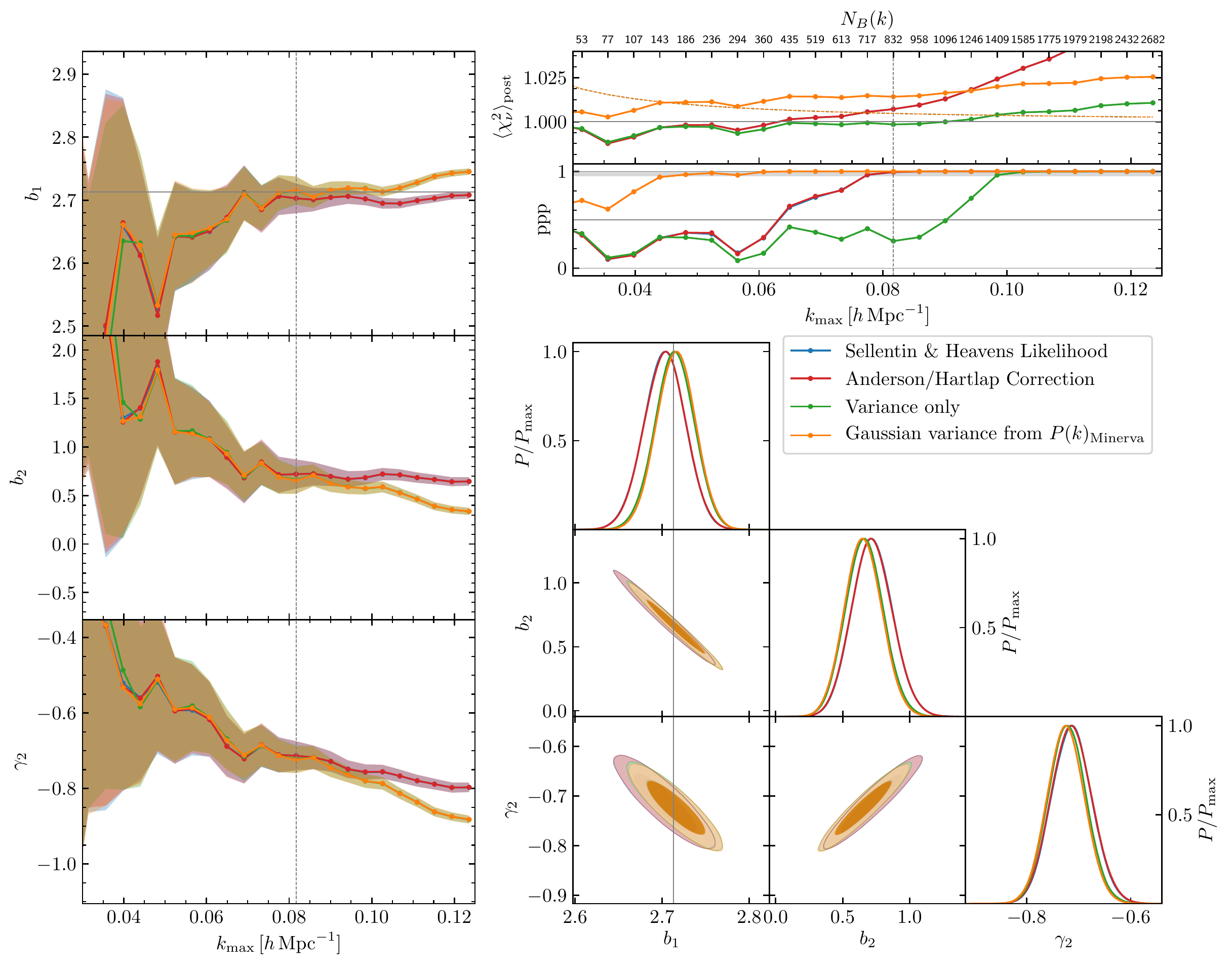}
    \caption{As in figure~\ref{fig:modelcomparisondk1} with $\Delta k=k_f$ but now comparing the different functional forms for the likelihood presented in section~\ref{sec:likshape} as well as the simple case of a Gaussian likelihood evaluated using only the diagonal part of the estimated covariance matrix. While the posteriors do not change in the range of validity of the model, the goodness-of-fit statistics give different results when using only the diagonal from the covariance estimated from the mocks or with the predicted Gaussian variance obtained in terms of the measured power spectrum.}
    \label{fig:likelihoodshapedk1}
\end{figure}

\begin{figure}[t]
    \centering
    \includegraphics[width=\textwidth]{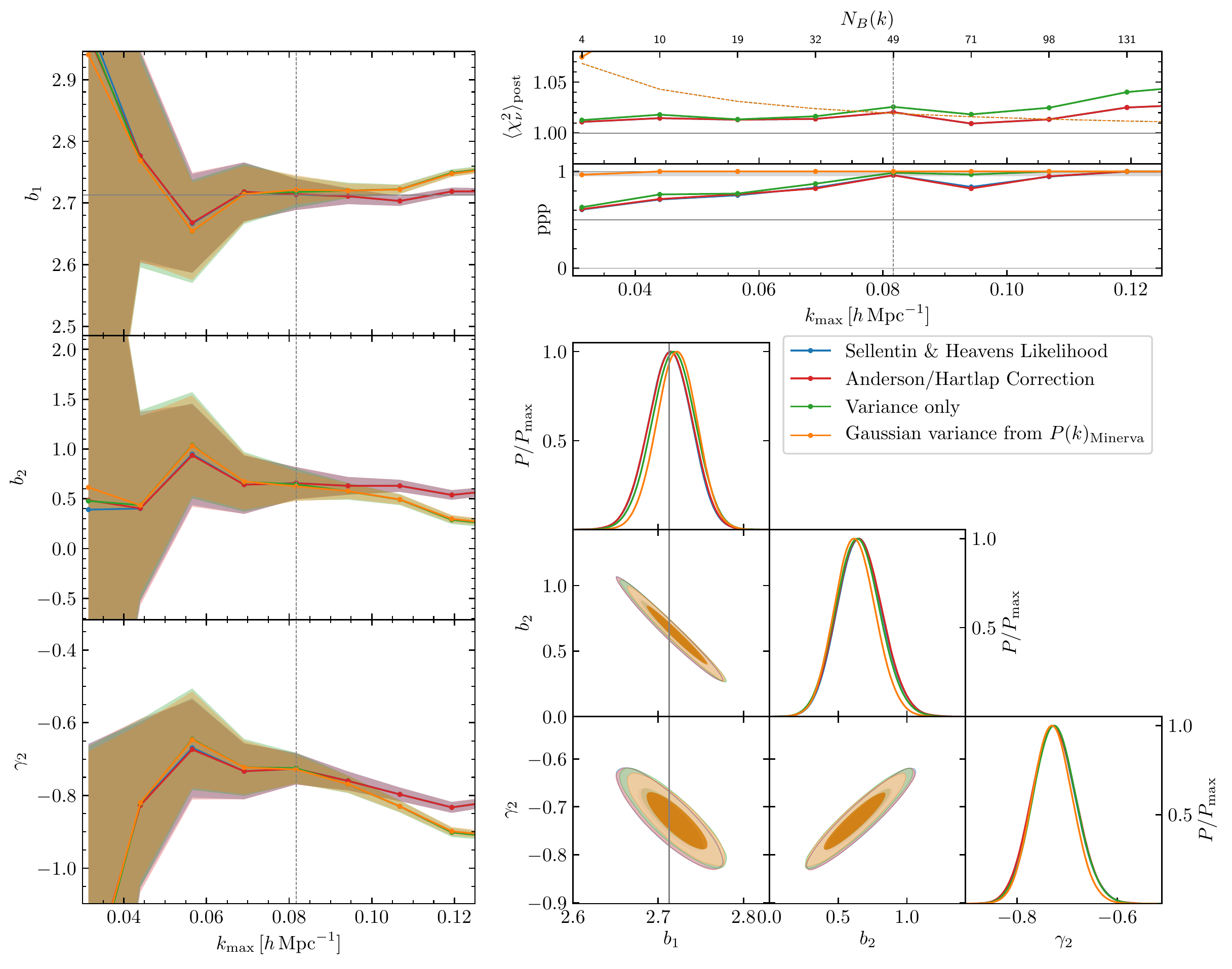}
    \caption{As in figure~\ref{fig:likelihoodshapedk1} but for $\Delta k= 3 k_f$.}
    \label{fig:likelihoodshapedk3}
\end{figure}

We now study how the shape of the likelihood function influences our results. To this purpose, we compare four options: i) the likelihood function introduced by \cite{SellentinHeavens2016} and presented in \eq{eq:SellentinLikelihood} that we have used to derive all the results presented so far; ii) a Gaussian likelihood combined with the unbiased estimate of the precision matrix given in \eq{eq:anderson} \cite{Anderson2003, HartlapSimonSchneider2007}; iii) the same as the previous case but considering only the diagonal part of the uncorrected covariance matrix for the data (\ie setting to zero all off-diagonal elements);
iv) the same as the previous case, but considering a Gaussian variance estimated using the full non-linear halo power spectrum as measured from the N-body simulations.
The corresponding fit results are displayed in figures~\ref{fig:likelihoodshapedk1} and \ref{fig:likelihoodshapedk3} for the data with $\Delta k = k_f$ and $3k_f$, respectively.
All cases refer to model $\mathcal{M}_3$ (fully averaged over the triangular bins). 
The first striking feature here is that, whenever the goodness of fit is acceptable, the posterior distributions for the model parameters are practically identical for all likelihood functions. This might be a consequence of the enormous compression involved in our exercise where we use tens of thousands of datapoints to measure only three parameters \footnote{Elena Sellentin, private communication.}. Also, the difference in the signal-to-noise ratio between the case of the full-covariance and its diagonal approximation shown in figure~\ref{fig:StoN} does not directly reflect in a noticeable difference in the final constraints.
Our results are obtained from measurements in periodic boxes at very large scales. Neglecting covariances is likely to produce larger biases in the presence of a window function and extending the analysis to more non-linear scales, when finite-volume effects and non-Gaussianity provide larger contributions to all off-diagonal elements of the covariance matrix \cite{SefusattiEtal2006}. 

Similarly, the goodness-of-fit statistics derived from the two likelihood functions that account for the off-diagonal covariances coincide almost perfectly. On the other hand, some deviations are noticeable when only the diagonal variances are considered. This approximation leads to rather optimistic estimates of the goodness of fit as a function of $\kmax$ for $\Delta k=k_f$ and to slightly pessimistic ones for $\Delta k=3k_f$ when only the diagonal part of the full estimated covariance is considered. On the other hand, in the case of the Gaussian variance prediction obtained from the measured power spectrum, the goodness of fit becomes even worse.
It is not easy to identify the precise origin of these effects, but it is important to stress that
the variance for the wider bins collects contributions from off-diagonal elements of the covariance matrix for the narrower bins.
This probably reduces, to some extent, the difference from the results obtained with the full covariance.

%%%%%%%%%%%%%%%%%%%%%%%%%%%%%%%%%%%%%%%%%%%%%%%%%%%%%%%%%%%%%%%%
%%%%%%%%%%%%%%%%%%%%%%%%%%%%%%%%%%%%%%%%%%%%%%%%%%%%%%%%%%%%%%%%
%%%%%%%%%%%%%%%%%%%%%%%%%%%%%%%%%%%%%%%%%%%%%%%%%%%%%%%%%%%%%%%%
\section{Conclusions}
\label{sec:conclusions}

In this work, we discuss how to fit the real-space halo bispectrum at large scales ($k \lesssim 0.1 \kMpc$) with a likelihood-based method. We consider dark-matter halos with a minimum mass of $10^{13}\Ms$ at redshift $z=1$, extracted from a large set of about three-hundred N-body simulations corresponding to a total volume of 1000$\cGpc$. This is much larger than the characteristic volume of current and forthcoming galaxy surveys: for instance, it is 100 times larger than the typical connected region that will be covered by the Euclid spectroscopic survey in a single redshift bin of, say, $\Delta z=0.1$ \cite{YankelevichPorciani2019}. Therefore, a model that fits our data well will generate systematic deviations below 10\% of the statistical errors expected for experiments such as Euclid or DESI. We precisely estimate the full, non-linear covariance for our data from an even larger set of 10,000 halo mock catalogs obtained with the Lagrangian PT-based {\pin} code. Statistical errors in the precision matrix are accounted for at the likelihood level with two different methods. 
We also pay particular attention to reduce any systematic error on the bispectrum covariance by carefully selecting the mass threshold for the mock halo catalogs. This provides a percent-level match to the N-body power spectrum that controls the main Gaussian contribution to the error budget. 
We consider a tree-level perturbative model for the halo bispectrum that allows quick numerical evaluations and therefore provides us with two advantages: MCMC chains run fast and binned theoretical predictions can be computed exactly. Our benchmark model depends on  two local bias parameters, $b_1$ and $b_2$, plus the quadratic tidal-field parameter $\gamma_2$ and assumes Poissonian shot noise. 

With these tools at our disposal, we study in the first place the goodness of fit. This issue is usually neglected in the literature, mainly due to the difficulty in obtaining an accurate and precise estimate of the covariance matrix. We determine the goodness of fit by means of the ppp introduced in section~\ref{sec:ppp}. To connect it with a more familiar diagnostic test, we also compute the posterior average of the reduced $\chi^2$, which gives consistent results. 
We find that the tree-level model provides a good fit to our data up to $0.08$-0.10$\kMpc$ depending on the binning of the Fourier modes adopted for the bispectrum measurements, that is $\Delta k= s k_f$ with $s=1$, 2 or 3. The deterioration of the goodness of fit for larger values of $k$ does not necessarily imply that the model needs higher-order corrections as it might instead reflect that our estimate of the covariance matrix for the bispectrum becomes less accurate.
The posterior distributions for the bias parameters are largely consistent across the different measurements. When plotted as a function of $\kmax$, they begin to show a marked scale dependence that, in conjunction with unusually small errors, indicates a failure for the model for scales $\kmax\gtrsim 0.09\kMpc$, consistent with the results for the goodness of fit.

We investigate several variations of our benchmark model by considering, for instance, either non-Poissonian shot noise or some specific relations between the bias parameters. 
We then apply Bayesian model selection to determine the optimal number of parameters that are needed to describe the numerical data without overfitting. In particular, our comparison is based on the BF (computed through the Savage-Dickey density ratio for nested theoretical models) and the DIC, both introduced in section~\ref{sec:MS}. Our results can be summarised as follows.
\begin{enumerate}
    \item The bispectrum data do not support the introduction of non-Poissonian shot-noise corrections. The BF strongly favours our benchmark model with respect to more elaborated models including additional shot-noise parameters, while the DIC is indecisive in this respect.
    \item Model-selection diagnostics clearly favour theoretical descriptions including a tidal bias term over a local Eulerian bias expansion with $\gamma_2=0$.
    \item A local Lagrangian bias expansions in which $\gamma_2=-\frac{2}{7}(b_1-1)$ is also disfavoured by the data with respect to models in which $\gamma_2$ is free to vary. On the other hand, using a fit from the literature to relate $\tilde{b}_2$ and $b_1$ (while keeping $\gamma_2$ free) gives a two-parameter model that is equivalent (in terms of DIC) to our benchmark description and leads to substantially smaller uncertainties for $b_1, b_2$ and $\gamma_2$. In a sense, using this fit is equivalent to combine the bispectrum with other data.
\end{enumerate}
Some of these conclusions confirm what other authors have found with different methods. 
We find remarkable that model-selection techniques applied to the bispectrum end up preferring exactly those models that are supported by other independent studies even when no difference can be noted in terms of simple goodness-of-fit diagnostics.
We envision that model selection will play an important role in the future as the number of parameters controlling loop models for the power spectrum and the bispectrum in redshift space will become particularly large with significant degeneracies among them  \cite{KimPorciani2019}.
Varying the cosmological parameters in order to analyze data from actual surveys will also introduce additional complications. We intend to address these issues in our future work.

Bispectrum measurements are invariably performed within finite bins collecting similar triangles of wavenumbers. Averaging the theoretical predictions over the same bins can be expensive if the model contains loop corrections. For this reason, we test several methods to reduce the number of model evaluations needed in order to fit binned data. We find that a single evaluation per bin (corresponding to a suitably defined effective triangle) generates only very minor systematic errors for $\Delta k=3\,k_f$.
On the other hand, naively using a triangle with sides corresponding to the bin centers leads to distinctly incorrect posteriors for the parameters of the fit.
 
As a final test, we consider different functional forms for the likelihood distribution. Two of them have been designed to account for statistical errors that plague the estimation of covariance matrices from mock catalogs. The first is obtained by marginalising over the unknown covariance matrix conditioned to its estimate \cite{SellentinHeavens2016}, the second simply combines a Gaussian likelihood with a re-scaled precision matrix \citep{HartlapEtal2009}. Lastly, as done in many previous works, we use a Gaussian likelihood and only consider either the diagonal part of the estimated covariance or the Gaussian variance prediction obtained in terms of the measured halo power spectrum from the N-body simulations. 
In all cases, we find essentially the same posterior distribution for the model parameters.
However, estimates of the goodness of fit get artificially inflated by neglecting covariances when fitting data within bins of $\Delta k = k_f$ (note that this does not happen for the wider bins).
No differences, instead, are noticeable between the two approaches that correct for the noise in the covariance matrix.
Based on this, we conclude that, for measurements in periodic boxes, inference performed considering only the variance leads to trustable posteriors while the range of scales over which a model provides an acceptable description of the data should be determined using the full covariance. We expect, however, that neglecting off-diagonal elements could not represent a viable option in an actual galaxy survey. In fact, window-function effects combined with non-Gaussianities in the galaxy distribution should generate significant contributions to the off-diagonal covariance terms.

As already mentioned, this paper should be regarded as the first step towards establishing a solid inference method for the galaxy bispectrum. By taking advantage of a reliable estimation of the covariance matrix for the halo bispectrum,  we have measured the influence of several, often overlooked, details.
In our future work, we will investigate other key methodological aspects. 
Among them, we will consider theoretical models that include loop corrections and test intrinsically non-Gaussian likelihood functions. Our ultimate goal is to set up a robust pipeline for extracting cosmological information from the joint analysis of the galaxy power spectrum and the bispectrum.

\acknowledgments

We are particularly grateful to Claudio Dalla Vecchia for granting us access to the Minerva simulations. We would like to thank Davit Alkhanishvili, Martin Crocce, Vincent Desjacques, Alexander Eggemeier, Dimitry Ginzburg, Titouan Lazeyras, Andrea Pezzotta, Roman Scoccimarro, Elena Sellentin and Victoria Yankelevich for valuable discussions.
The Minerva simulations have been performed and analysed on the Hydra and Euclid clusters at the Max Planck Computing and Data Facility (MPCDF) in Garching. The {\pin} mocks were run on the GALILEO cluster at CINECA, thanks to an agreement with the University of Trieste.
This research was supported by the Munich Institute for Astro- and Particle Physics (MIAPP) of the DFG cluster of excellence ``Origin and Structure of the Universe'' where part of the final draft was completed.
AO, ES, and PM are partially supported by the INFN INDARK PD51 grant.
ES and PM acknowledge support from PRIN MIUR 2015 Cosmology and Fundamental Physics: illuminating the Dark Universe with Euclid.

%%%%%%%%%%%%%%%%%%%%%%%%%%%%%%%%%%%%%%%%%%%%%%%%%%%%%%%%%%%%%%%%
%%%%%%%%%%%%%%%%%%%%%%%%%%%%%%%%%%%%%%%%%%%%%%%%%%%%%%%%%%%%%%%%
%%%%%%%%%%%%%%%%%%%%%%%%%%%%%%%%%%%%%%%%%%%%%%%%%%%%%%%%%%%%%%%%

\setlength{\bibsep}{2pt plus 0.5ex}
\bibliographystyle{JHEP}
\bibliography{cosmologia}

\end{document}